\documentclass[12pt,twoside]{article}
\usepackage{amssymb} 
\usepackage{amsmath} 
\usepackage{graphicx}
\usepackage{hyperref} 
\usepackage{lscape} 
\usepackage{xcolor} 
\begin{document} 
\noindent   	 
{\Large\bf A new matrix representation of the Maxwell equations based on the Riemann-Silberstein-Weber vector for a linear inhomogeneous medium} 

\bigskip 

\noindent
{\large\bf Sameen Ahmed Khan$^{a}\,$\footnote{Corresponding author, 
{\em email}:~rohelakhan@yahoo.com,~ORCID~ID: 
\url{https://orcid.org/0000-0003-1264-2302}}, 
Ramaswamy Jagannathan$^{b}\,$\footnote{Retired Faculty,  
{\em email}:~jagan@imsc.res.in, ORCID~ID:  \url{https://orcid.org/0000-0003-2968-2044}}} 

\medskip 

\noindent 
$^{a}${\it Department of Mathematics and Sciences, College of Arts and Applied Sciences, Dhofar University, Salalah, Sultanate of Oman}. \\   
$^{b}${\it The Institute of Mathematical Sciences, Central Institutes of Technology Campus, Tharamani, Chennai, India}.   

\vspace{1cm}  	 

\noindent 
ABSTRACT 

\smallskip 

\noindent  
We derive a new eight dimensional matrix representation of the Maxwell equations 
for a linear homogeneous medium and extend it to the case of a linear inhomogneous medium.  This derivation starts {\em ab initio} with the Maxwell equations and uses arguments based on the algebra of the Pauli matrices.  This process leads automatically to the matrix representation based on the Riemann-Silberstein-Weber (RSW) vector.  The new representation for the homogeneous medium is a direct sum 
of four Pauli matrix blocks.  This aspect of the new representation should make it suitable for studying the propagation of electromagnetic waves in a linear inhomogeneous medium adopting the techniques of quantum mechanics treating the inhomogeneity as a perturbation.  The new representation is used to rederive the Mukunda-Simon-Sudarshan matrix substitution rule for transition from the Helmholtz scalar wave optics to the Maxwell vector wave optics.  

\bigskip 


\newpage  

\section{Introduction} 

Representing the Maxwell equations in a matrix form, or essentially in a Dirac-like form, mostly for vacuum, has a long history starting with the works of Laporte and Uhlenbeck, Oppenheimer, Majorana, and others (see Laporte and Uhlenbeck, 1931;  Oppenheimer, 1931; Moliere, 1950; Good, 1957; Moses, 1959; Mignani et al., 1974;  Edmonds, 1975; Bialynicki-Birula, 1994, 1996a,b; Esposito, 1998; Bogush et al., 2009; Kisel et al., 2011; Bialynicki-Birula and Bialynicki-Birula, 2013; Barnett, 2014; Belkovich and Kogan, 2016; Kulyabov, 2016; Kulyabov et al., 2017; Kiesslinga and Tahvildar-Zadehb, 2018; Livadiotis, 2018; Jestadt et al., 2019; Sebens, 2019; Mingjie et al., 2020; Khan, 2016a; Korotkova and Testorf, 2023; (see, {\em e.g.}, Bialynicki-Birula, 1996b; Bogush et al., 2009; Kisel et al., 2011; for a comprehensive guide to the literature on the topic).  These numerous matrix representations of the Maxwell equations, mostly six dimensional and based on the Riemann-Silberstein-Weber (RSW) vector, have been studied from different perspectives such as symmetries, mathematical correspondence with the spinor Dirac equation, etc., and are not primarily intended for applications to optics.  With applications to optics in mind, an eight dimensional matrix representation of the Maxwell equations for an inhomogneous medium was derived by one of us (Khan, 2005) heuristically, based on the RSW vector, following (Bialynicki-Birula, 1994, 1996a,b).  The representation in (Khan, 2005) has been found to be useful in studying several problems in electromagnetic theory (Khan, 2008, 2010, 2014, 2016, 2017a,b, 2023a,b, 2024; Mehrafarin and Balajany, 2010; Ram et al., 2021; Vahala et al., 2020a,b,c, 2021a,b, 2022a,b; Jin et al., 2023).  The representation in 
(Khan, 2005) has been used to study the optics of the Maxwell vector wave beams (Khan, 2008, 2010, 2014, 2016a, 2017a,b) based on the analogy with the quantum mechanics of charged particle beam optics (Jagannathan et al., 1989; Jagannathan, 1990, 1999, 2002, 2004; Khan and Jagannathan, 1995, 2021, 2024a,b; Conte et al., 1996; Khan, 1997, 1999, 2002; Jagannathan and Khan, 1996, 2019; Hawkes, 2020; Hawkes and Kasper, 2022).  The RSW vector is commonly known in the literature as the Riemann-Silberstein vector.  Following (Ram et al., 2021; Vahala et al., 2022a,b) we call it the Riemann-Silberstein-Weber vector in view of the significance of the contributions to it by Weber (see Kiesslinga and Tahvildar-Zadehb, 2018; Sebens, 2019; for details).  

Here, we derive first a new eight dimensional matrix representation of the Maxwell equations for a linear homogeneous medium starting {\it ab initio} with the Maxwell equations following (Bocker and Frieden, 1993, 2018) wherein an eight dimensional matrix representation of the Maxwell equations for vacuum has been studied similarly.  To this end, we use the arguments based on the matrix representations of the Pauli algebra.  This process leads automatically to the eight dimensional matrix representation of the Maxwell equations for a linear homogeneous medium based on the RSW vector.  In this sense, the RSW vector turns out to be the natural choice for building a matrix representation of the Maxwell equations.  This representation for the homogeneous medium is a direct sum of four Pauli matrix blocks (see Appendix A for the definition of the direct sum of matrices).  Then, we extend the new representation to the case of a linear inhomogeneous medium.  

The new eight dimensional matrix representation of the Maxwell equations for a linear inhomogeneous medium reduces to the representation for the linear homogeneous medium when there is no inhomogeneity.  The representation for the homogeneous medium is the direct sum of four Pauli matrix blocks.  Pauli matrices have a very simple algebra. This aspect of the new representation should make it suitable for developing perturbation methods for studying the propagation of electromagnetic waves in a linear inhomogeneous medium adopting the techniques of quantum mechanics treating the inhomogeneity as a perturbation.  We also show that the charge continuity equation, inhomogeneous wave equations for the electric and magnetic fields, and the generalized Helmholtz equation, emerge from the new eight dimensional matrix representation of the Maxwell equations.  We shall use the complex notation for the electromagnetic waves as is customary in optics (see, 
{\em e.g.}, Born and Wolf, 1999). 

It has been well recognized that scalar wave optical description of electromagnetic beams of finite transverse size is inconsistent with the Maxwell equations 
(Lax et al., 1975).  Addressing this problem, a systematic procedure for transition from the scalar wave optics to the vector wave optics consistent with the Maxwell equations has been derived from a group theoretical analysis of the Maxwell theory (Sudarshan et al., 1983; Mukunda et al., 1983).  This procedure has been used to construct vector wave beams consistent with the Maxwell equations (Mukunda et al., 1985a,b; Simon et al., 1986, 1987).  We expect a matrix representation of the Maxwell equations to provide an alternative approach to the problem.  Already, the Mukunda-Simon-Sudarshan (MSS) matrix substitution rule (Mukunda et al., 1983) for transition from the Helmholtz scalar wave optics to the Maxwell vector wave optics has been rederived (Khan, 2016b) using a four dimensional approximation of the eight dimensional matrix representation of the Maxwell equations given in (Khan, 2005).  We present here an alternative derivation of the MSS substitution rule based on the new eight dimensional matrix representation without any approximation.     
 
\section{The new matrix representation of the Maxwell equations for a linear homogeneous medium}  
 
Let us consider the Maxwell equations for a medium, in presence of charges and currents,  
\begin{eqnarray}  
\frac{\partial\mbox{\boldmath $D$}(\mbox{\boldmath $r$},t)}{\partial t} 
& = & \mbox{\boldmath $\nabla$} \times 
\mbox{\boldmath $H$}(\mbox{\boldmath $r$},t)
-\mbox{\boldmath $J$}(\mbox{\boldmath $r$},t),  
\nonumber \\ 
\mbox{\boldmath $\nabla$} \cdot \mbox{\boldmath $D$}(\mbox{\boldmath $r$},t) 
& = & \rho(\mbox{\boldmath $r$},t),  
\nonumber \\  
\frac{\partial\mbox{\boldmath $B$}(\mbox{\boldmath $r$},t)}{\partial t} 
& = & -\mbox{\boldmath $\nabla$} \times 
\mbox{\boldmath $E$}(\mbox{\boldmath $r$},t),  
\nonumber \\  
\mbox{\boldmath $\nabla$} \cdot \mbox{\boldmath $B$}(\mbox{\boldmath $r$},t) 
& = & 0.   
\label{Maxwell}
\end{eqnarray} 
We assume the medium to be linear such that  
\begin{equation} 
\mbox{\boldmath $D$}(\mbox{\boldmath $r$},t) 
= \epsilon(\mbox{\boldmath $r$},t)\mbox{\boldmath $E$}(\mbox{\boldmath $r$},t),  \qquad  
\mbox{\boldmath $B$}(\mbox{\boldmath $r$},t) 
= \mu(\mbox{\boldmath $r$},t)\mbox{\boldmath $H$}(\mbox{\boldmath $r$},t).  
\end{equation} 
Here, $\epsilon(\mbox{\boldmath $r$},t)$ and $\mu(\mbox{\boldmath $r$},t)$ are the permittivity and permeability of the medium, respectively.  Then, 
\begin{equation}
v(\mbox{\boldmath $r$},t) 
= \frac{1}{\sqrt{\epsilon(\mbox{\boldmath $r$},t)\mu(\mbox{\boldmath $r$},t)}},  
\end{equation}  
is the speed of light in the medium and the speed of light in vacuum is   
$c = 1/{\sqrt{\epsilon_0\mu_0}}$ with $\epsilon_0$ and $\mu_0$, respectively, as 
the permittivity and permeability of vacuum.  The refractive index of the medium is given by 
\begin{equation} 
n(\mbox{\boldmath $r$},t) 
= \frac{c}{v(\mbox{\boldmath $r$},t)} 
= c\sqrt{\epsilon(\mbox{\boldmath $r$},t)\mu(\mbox{\boldmath $r$},t)},  
\end{equation} 
and 
\begin{equation}   
\eta(\mbox{\boldmath $r$},t) 
= \sqrt{\frac{\mu(\mbox{\boldmath $r$},t)}{\epsilon(\mbox{\boldmath $r$},t)}}   
\end{equation} 
is the impedance of the medium.  The fields $\mbox{\boldmath $E$}$, 
$\mbox{\boldmath $D$}$, $\mbox{\boldmath $B$}$, and $\mbox{\boldmath $H$}$ and the quantities $\epsilon$, $\mu$, $v$, $n$, and $\eta$ are local and instantaneous.  Hereafter, we shall not indicate explicitly the space-time $(\mbox{\boldmath $r$},t)$ dependence of $\mbox{\boldmath $E$}$, $\mbox{\boldmath $D$}$, 
$\mbox{\boldmath $B$}$, $\mbox{\boldmath $H$}$, $\epsilon$, $\mu$, $v$, $n$, and $\eta$, unless required.  Rewrite Eq. (\ref{Maxwell}) as 
\begin{eqnarray} 
\frac{\partial (\sqrt{\epsilon} 
(\sqrt{\epsilon}\mbox{\boldmath $E$}))}{\partial t} 
& = & \mbox{\boldmath $\nabla$} \times \left( \frac{1}{\sqrt{\mu}} 
\left( \frac{\mbox{\boldmath $B$}}{\sqrt{\mu}} \right) \right) 
- \mbox{\boldmath $J$},  
\nonumber \\  
\mbox{\boldmath $\nabla$} \cdot (\sqrt{\epsilon} 
(\sqrt{\epsilon}\mbox{\boldmath $E$})) 
& = & \rho,  
\nonumber \\  
\frac{\partial}{\partial t} \left( \sqrt{\mu} \left( \frac{\mbox{\boldmath $B$}}
{\sqrt{\mu}} \right)\right)  
& = & -\mbox{\boldmath $\nabla$} \times \left( \frac{1}{\sqrt{\epsilon}} 
\left( \sqrt{\epsilon} \mbox{\boldmath $E$} \right)\right),  
\nonumber \\   
\mbox{\boldmath $\nabla$} \cdot \left( \sqrt{\mu}
\left( \frac{\mbox{\boldmath$B$}}{\sqrt{\mu}} \right)\right) 
& = & 0.    
\label{MEqepsmu} 
\end{eqnarray}  
For a homogeneous time-independent medium with constant $\epsilon$ and $\mu$ 
Eq. (\ref{MEqepsmu}) becomes 
\begin{eqnarray} 
\frac{\partial(\sqrt{\epsilon}\mbox{\boldmath $E$})}{\partial t} 
& = & v\mbox{\boldmath $\nabla$} \times 
\left(\frac{\mbox{\boldmath $B$}}{\sqrt{\mu}}\right) 
- \frac{\mbox{\boldmath $J$}}{\sqrt{\epsilon}},  
\nonumber \\ 
\mbox{\boldmath $\nabla$} \cdot (\sqrt{\epsilon}\mbox{\boldmath $E$})    
& = & \frac{\rho}{\sqrt{\epsilon}},  
\nonumber \\     
\frac{\partial}{\partial t}\left(\frac{\mbox{\boldmath $B$}}{\sqrt{\mu}}\right)  
& = & -v\mbox{\boldmath $\nabla$} \times 
\left(\sqrt{\epsilon}\mbox{\boldmath $E$}\right),  
\nonumber \\ 
\mbox{\boldmath $\nabla$} \cdot 
\left(\frac{\mbox{\boldmath $B$}}{\sqrt{\mu}}\right) 
& = & 0,  
\label{MEqconstepsmu}
\end{eqnarray} 
where $v = c/n$ is constant.  Following (Bocker and Frieden, 1993, 2018), we write Eq. (\ref{MEqconstepsmu}) in a matrix form as  
\begin{subequations} 
\label{dFdtMFJ}  
\begin{equation} 
\frac{\partial \mathcal{F}}{\partial t} 
= v{\mathcal{M}}_0 {\mathcal{F}} - {\mathcal{J}},  
\label{dFdtMFJ-a}
\end{equation} 
\begin{equation} 
{\mathcal{F}} 
= \frac{1}{\sqrt{2}} 
\left(\begin{array}{c} 
\sqrt{\epsilon} E_x \\ \sqrt{\epsilon} E_y \\ \sqrt{\epsilon} E_z 
\\ 0 \\ 
\frac{1}{\sqrt{\mu}} B_x \\ \frac{1}{\sqrt{\mu}} B_y \\ \frac{1}{\sqrt{\mu}} B_z 
\\ 0 
\end{array}\right),  \qquad 
{\mathcal{J}} 
= \frac{1}{\sqrt{2\epsilon}} 
\left(\begin{array}{c} 
J_x \\ J_y \\ J_z \\ 0 \\ 0 \\ 0 \\ 0 \\ - v \rho          
\end{array}\right),    
\label{dFdtMFJ-b}
\end{equation} 
\begin{equation}         
{\mathcal{M}}_0 
= \left(\begin{array}{rrrrrrrr}
0 &  0 &  0 &  0 & 0 & -\partial_z & \partial_y & -\partial_x \\ 
0 &  0 &  0 &  0 & \partial_z & 0 & -\partial_x & -\partial_y \\ 
0 &  0 &  0 &  0 & -\partial_y & \partial_x &  0 & -\partial_z \\ 
0 &  0 &  0 &  0 & \partial_x & \partial_y & \partial_z &  0 \\ 
0 & \partial_z & -\partial_y & \partial_x & 0 & 0 &  0 &  0 \\ 
-\partial_z &  0 & \partial_x & \partial_y & 0 & 0 &  0 &  0 \\ 
\partial_y & -\partial_x &  0 & \partial_z & 0 & 0 &  0 &  0 \\ 
-\partial_x & -\partial_y &  -\partial_z &  0 & 0 & 0 &  0 &  0 
\end{array}\right), 
\label{dFdtMFJ-c}  
\end{equation} 
\end{subequations} 
where $\partial_x = {\partial}/{\partial x}$, 
$\partial_y = {\partial}/{\partial y}$, $\partial_z = {\partial}/{\partial z}$.  
Note that the entries of the 4th and 8th rows and columns of the matrix 
${\mathcal{M}}_0$ are not uniquely determined in Eq. (\ref{dFdtMFJ}).   These entries have been fixed by considerations of symmetry, following (Bocker and Frieden, 1993, 2018), such that if we write 
\begin{equation}
{\mathcal{M}}_0 
= {\mathcal{M}}_{0x} \partial_x + {\mathcal{M}}_{0y} \partial_y 
+ {\mathcal{M}}_{0z} \partial_z  
= {\mbox{\boldmath $\mathcal{M}$}}_0 \cdot {\mbox{\boldmath $\nabla$}},     
\label{mathcalM0}
\end{equation} 
then, the matrix coefficients of $\partial_x$, $\partial_y$, and $\partial_z$ are Hermitian, {\em i.e.}, 
\begin{subequations} 
\label{mathcalM0xyz} 
\begin{equation}     
{\mathcal{M}}_{0x} 
= \left(\begin{array}{rrrrrrrr}
0 &  0 &  0 &  0 & 0 & 0 &  0 & -1 \\ 
0 &  0 &  0 &  0 & 0 & 0 & -1 &  0 \\ 
0 &  0 &  0 &  0 & 0 & 1 &  0 &  0 \\ 
0 &  0 &  0 &  0 & 1 & 0 &  0 &  0 \\ 
0 &  0 &  0 &  1 & 0 & 0 &  0 &  0 \\ 
0 &  0 &  1 &  0 & 0 & 0 &  0 &  0 \\ 
0 & -1 &  0 &  0 & 0 & 0 &  0 &  0 \\ 
-1 &  0 &  0 &  0 & 0 & 0 &  0 &  0 
\end{array}\right)  
= {\mathcal{M}}_{0x}^{\dagger}, 
\end{equation} 
\begin{equation}   
{\mathcal{M}}_{0y} 
= \left(\begin{array}{rrrrrrrr}
0 &  0 &  0 &  0 &  0 & 0 &  1 &  0 \\ 
0 &  0 &  0 &  0 &  0 & 0 &  0 & -1 \\ 
0 &  0 &  0 &  0 & -1 & 0 &  0 &  0 \\ 
0 &  0 &  0 &  0 &  0 & 1 &  0 &  0 \\ 
0 &  0 & -1 &  0 &  0 & 0 &  0 &  0 \\ 
0 &  0 &  0 &  1 &  0 & 0 &  0 &  0 \\ 
1 &  0 &  0 &  0 &  0 & 0 &  0 &  0 \\ 
0 & -1 &  0 &  0 &  0 & 0 &  0 &  0 
\end{array}\right) 
= {\mathcal{M}}_{0y}^{\dagger}, 
\end{equation} 
\begin{equation} 
{\mathcal{M}}_{0z}
= \left(\begin{array}{rrrrrrrr}
0 &  0 &  0 &  0 &  0 & -1 &  0 &  0 \\ 
0 &  0 &  0 &  0 &  1 &  0 &  0 &  0 \\ 
0 &  0 &  0 &  0 &  0 &  0 &  0 & -1 \\ 
0 &  0 &  0 &  0 &  0 &  0 &  1 &  0 \\ 
0 &  1 &  0 &  0 &  0 &  0 &  0 &  0 \\ 
-1 &  0 &  0 &  0 &  0 &  0 &  0 &  0 \\ 
0 &  0 &  0 &  1 &  0 &  0 &  0 &  0 \\ 
0 &  0 & -1 &  0 &  0 &  0 &  0 &  0 
\end{array}\right)  
= {\mathcal{M}}_{0z}^{\dagger}.    
\end{equation} 
\end{subequations}  
where ${X}^{\dagger}$ denotes the Hermitian conjugate of $X$.  

It should be noted that 
\begin{eqnarray} 
{\mathcal{M}}_{0j}^2 
& = & {\mbox{\boldmath $1$}}_8, \qquad \mbox{for}\ j = x,y,z, 
\nonumber \\   
{\mathcal{M}}_{0j}{\mathcal{M}}_{0k} 
& = & -{\mathcal{M}}_{0k}{\mathcal{M}}_{0j}, \qquad 
\mbox{for}\ j \neq k, \quad  j,k = x,y,z.   
\label{Malgebra} 
\end{eqnarray} 
Here, $\mbox{\boldmath $1$}_8$ is the eight dimensional identity matrix, 
\begin{equation}
{\mbox{\boldmath $1$}}_8 
= \left(\begin{array}{cccccccc}
1 & 0 & 0 & 0 & 0 & 0 & 0 & 0 \\ 
0 & 1 & 0 & 0 & 0 & 0 & 0 & 0 \\ 
0 & 0 & 1 & 0 & 0 & 0 & 0 & 0 \\ 
0 & 0 & 0 & 1 & 0 & 0 & 0 & 0 \\
0 & 0 & 0 & 0 & 1 & 0 & 0 & 0 \\
0 & 0 & 0 & 0 & 0 & 1 & 0 & 0 \\
0 & 0 & 0 & 0 & 0 & 0 & 1 & 0 \\
0 & 0 & 0 & 0 & 0 & 0 & 0 & 1  
\end{array}\right).  
\end{equation}
Similarly, later we shall write 
\begin{eqnarray}
{\mbox{\boldmath $1$}}_2 
& = & \left(\begin{array}{cc}
1 & 0 \\ 
0 & 1    
\end{array}\right),  \qquad 
{\mbox{\boldmath $1$}}_4 
= \left(\begin{array}{cccc}
1 & 0 & 0 & 0 \\ 
0 & 1 & 0 & 0 \\ 
0 & 0 & 1 & 0 \\ 
0 & 0 & 0 & 1   
\end{array}\right),     
\nonumber \\ 
{\mbox{\boldmath $0$}}_2 
& = & \left(\begin{array}{cc}
0 & 0 \\ 
0 & 0   
\end{array}\right),  \qquad 
{\mbox{\boldmath $0$}}_4 
= \left(\begin{array}{cccc}
0 & 0 & 0 & 0 \\ 
0 & 0 & 0 & 0 \\ 
0 & 0 & 0 & 0 \\ 
0 & 0 & 0 & 0   
\end{array}\right).   
\end{eqnarray} 
    
Divide Eq. (\ref{dFdtMFJ-a}) throughout by $v$ and rewrite it as 
\begin{equation} 
\left(\frac{1}{v}\frac{\partial}{\partial t} - \mathcal{M}_0 \right)\mathcal{F} 
= -\frac{1}{v}\mathcal{J}.  
\label{1byvdFdtMFJ} 
\end{equation} 
Note that ${\partial}/{\partial t}$ and $\mathcal{M}_0$ are commutative.  It follows from Eqs. (\ref{mathcalM0}) and (\ref{Malgebra}) that 
\begin{equation} 
\mathcal{M}_0^2 
= {\mbox{\boldmath $1$}}_8\nabla^2,  \qquad 
\nabla^2 = \left(\partial_x^2 + \partial_y^2 + \partial_z^2\right).   
\end{equation} 
Multiplying both sides of Eq. (\ref{1byvdFdtMFJ}) by 
$\left( (1/v)({\partial}/{\partial t}) + \mathcal{M}_0 \right)$ gives us 
\begin{equation} 
\left(\frac{1}{v^2}\frac{\partial^2}{\partial t^2} - \nabla^2 \right)\mathcal{F} 
= -\frac{1}{v}\left(\frac{1}{v}\frac{\partial}{\partial t} + \mathcal{M}_0\right)
\mathcal{J}.  
\end{equation} 
If we carry out the multiplication on the right hand side, we get the inhomogeneous wave equations 
\begin{eqnarray} 
\frac{1}{v^2}\frac{\partial^2 \mbox{\boldmath {$E$}}}{\partial t^2} 
- \nabla^2 \mbox{\boldmath {$E$}}  
& = & -\frac{1}{\epsilon}\mbox{\boldmath {$\nabla$}}\rho  
+ \mu\frac{\partial\mbox{\boldmath {$J$}}}{\partial t},  
\nonumber \\  
\frac{1}{v^2} \frac{\partial^2\mbox{\boldmath {$B$}}}{\partial t^2} 
- \nabla^2\mbox{\boldmath {$B$}} 
& = & \mu\left(\mbox{\boldmath {$\nabla$}} \times \mbox{\boldmath {$J$}}\right), 
\label{waveeqns} 
\end{eqnarray} 
and the charge continuity equation 
\begin{equation} 
\mbox{\boldmath {$\nabla$}} \cdot \mbox{\boldmath {$J$}}  
+ \frac{\partial \rho}{\partial t} = 0.     
\end{equation}   
Thus, fixing the non-unique entries of ${\mathcal{M}}_0$ such that ${\mathcal{M}}_{0x}$, ${\mathcal{M}}_{0y}$, and ${\mathcal{M}}_{0z}$ are Hermitian, makes Eq. (\ref{dFdtMFJ}) consistent with the inhomogneous wave equations for $\mbox{\boldmath $E$}$ and $\mbox{\boldmath $B$}$ and the charge continuity equation.  For a source-free homogeneous medium Eq. (\ref{waveeqns}) reduces to the Helmholtz equations 
\begin{equation}  
\frac{1}{v^2}\frac{\partial^2 \mbox{\boldmath {$E$}}}{\partial t^2} 
- \nabla^2 \mbox{\boldmath {$E$}} = 0,  \qquad 
\frac{1}{v^2}\frac{\partial^2 \mbox{\boldmath {$B$}}}{\partial t^2} 
- \nabla^2 \mbox{\boldmath {$B$}} = 0. 
\label{Helmholtz}
\end{equation} 

Hereafter, we shall use the direct product notation to write the matrices in a shorter form wherever suitable (see Appendix A for the definition of the direct product of matrices).  Then, in terms of the three Pauli matrices 
\begin{equation}
\sigma_x 
= \left(\begin{array}{cc} 
0 & 1 \\ 1 & 0 \end{array} \right), \quad 
\sigma_y 
= \left(\begin{array}{cr} 
0 & -i \\ i & 0 \end{array} \right), \quad   
\sigma_z 
= \left(\begin{array}{cr} 
1 & 0 \\ 0 & -1 \end{array} \right),    
\label{Pauli}
\end{equation}
we have 
\begin{eqnarray} 
{\mathcal{M}}_{0x}
& = & \left(\sigma_y\otimes\sigma_y\otimes\sigma_x\right), \nonumber \\  
{\mathcal{M}}_{0y} 
& = & -\left(\sigma_y\otimes\sigma_y\otimes\sigma_z\right), \nonumber \\    
{\mathcal{M}}_{0z}	
& = & \left(\sigma_y\otimes{\mbox{\boldmath $1$}}_2\otimes\sigma_y\right).  
\label{M0xyz}
\end{eqnarray} 
The three Pauli matrices are Hermitian, {\em i.e.}, $\sigma_j^\dagger = \sigma_j$, for $j = x,y,z$.  From this and Eq. (\ref{M0xyz}) it follows that ${\mathcal{M}}_{0x}$, ${\mathcal{M}}_{0y}$, and ${\mathcal{M}}_{0z}$ are Hermitian as we have found already in Eq. (\ref{mathcalM0xyz}).  We know that in quantum mechanics the canonical momentum operator $\widehat{\mbox{\boldmath $p$}} = -i\hbar{\mbox{\boldmath $\nabla$}}$ is Hermitian.  Hence it follows that 
${\mbox{\boldmath $\nabla$}} = (i/\hbar)\widehat{\mbox{\boldmath $p$}}$ is anti-Hermitian {\em i.e.}, ${\mbox{\boldmath $\nabla$}}^\dagger = 
-{\mbox{\boldmath $\nabla$}}$.  Then, we see from Eq. (\ref{mathcalM0}) that ${\mathcal{M}}_0^\dagger = -{\mathcal{M}}_0$, {\em i.e.}, ${\mathcal{M}}_0$ is anti-Hermitian.  As we shall see later, this anti-Hermitian nature of ${\mathcal{M}}_0$ leads to the unitary time evolution of ${\mathcal{F}}(t)$ and conservation of energy for a time-independent source-free homogeneous medium. 

The three Pauli matrices obey the algebra 
\begin{equation}  
\sigma_j^2 
= {\mbox{\boldmath $1$}}_2,\ \mbox{for}\ j = x,y,z, \quad 
\sigma_j\sigma_k 
= -\sigma_k\sigma_j,\ \mbox{for}\ j \neq k,\ \ j,k = x,y,z. 
\label{Paulialgebra} 
\end{equation} 
Comparing Eq. (\ref{Malgebra}) with Eq. (\ref{Paulialgebra}) shows that the set of matrices $\left\{{\mathcal{M}}_{0x}, {\mathcal{M}}_{0y}, {\mathcal{M}}_{0z}\right\}$ provides an eight dimensional faithful representations of the algebra of the Pauli matrices.  It is well known that there are only two inequivalent irreducible faithful  representations of the Pauli algebra: 
$\left\{ \sigma_x,\sigma_y,\sigma_z\right\}$ and $\left\{-\sigma_x,-\sigma_y,-\sigma_z\right\}$ (see, {\em e.g.}, Weinberger, 2008).  The latter is unitarily equivalent to 
$\left\{\sigma_x^*, \sigma_y^*, \sigma_z^* \right\}$ $=$ 
$\left\{\sigma_x, -\sigma_y, \sigma_z \right\}$, with $^*$ denoting the complex conjugate.  This is seen from the following observation:  
$\sigma_y\sigma_x\sigma_y^\dagger = -\sigma_x$,  
$\sigma_y\sigma_y^*\sigma_y^\dagger = -\sigma_y$,  
$\sigma_y\sigma_z\sigma_y^\dagger = -\sigma_z$.  Note that 
$\sigma_x\sigma_y\sigma_z = i{\mbox{\boldmath $1$}}_2$ and  $(-\sigma_x)(-\sigma_y)(-\sigma_z) = -i{\mbox{\boldmath $1$}}_2$ characterizing their inequivalence. Thus, the eight dimensional faithful representation of the Pauli algebra provided by $\left\{{\mathcal{M}}_{0x},{\mathcal{M}}_{0y},
{\mathcal{M}}_{0z}\right\}$ should be completely reducible to a direct sum of four irreducible two dimensional representations.  This will lead to a reduction of ${\mathcal{M}_0}$ into a direct sum of four two dimensional blocks.  We shall find this reduced form of ${\mathcal{M}_0}$ by explicit construction.   

Let us observe that all the three matrices $\left\{{\mathcal{M}}_{0x}, {\mathcal{M}}_{0y}, {\mathcal{M}}_{0z}\right\}$ contain $\sigma_y$ as the first factor in the direct product representation in Eq. (\ref{M0xyz}).  Thus, if we perform a similarity transformation of all these three  matrices by a matrix which diagonalizes $\sigma_y$ and leaves the other factors unchanged in the direct product representation in Eq. (\ref{M0xyz}), then, all these three matrices will be block diagonalized as a direct sum of two four dimensional blocks.  To this end, we proceed as follows.  Consider the matrix 
\begin{equation} 
\tau 
= \frac{1}{\sqrt{2}}\left( {\mbox{\boldmath $1$}}_2 + i\sigma_x \right) 
= \frac{1}{\sqrt{2}}
\left(\begin{array}{cc} 
1 & i \\ i & 1 
\end{array}\right).    
\end{equation} 
It is unitary, {\em i.e.}, 
$\tau^\dagger\tau = \tau\tau^\dagger = {\mbox{\boldmath $1$}}_2$, and it diagonalizes $\sigma_y$ such that 
\begin{equation} 
\tau\sigma_y\tau^\dagger = -\sigma_z.   
\label{tausigmatau}
\end{equation} 
Let 
\begin{eqnarray} 
{\mathcal{T}} 
& = & \tau\otimes{\mbox{\boldmath $1$}}_4  
\nonumber \\ 
& = & \frac{1}{\sqrt{2}} 
\left(\begin{array}{cccccccc} 
1 & 0 & 0 & 0 & i & 0 & 0 & 0 \\ 
0 & 1 & 0 & 0 & 0 & i & 0 & 0 \\
0 & 0 & 1 & 0 & 0 & 0 & i & 0 \\ 
0 & 0 & 0 & 1 & 0 & 0 & 0 & i \\ 
i & 0 & 0 & 0 & 1 & 0 & 0 & 0 \\ 
0 & i & 0 & 0 & 0 & 1 & 0 & 0 \\
0 & 0 & i & 0 & 0 & 0 & 1 & 0 \\ 
0 & 0 & 0 & i & 0 & 0 & 0 & 1  
\end{array} \right).  
\label{Tau} 
\end{eqnarray} 
The matrix $\mathcal{T}$ is unitary: 
$\mathcal{T}\mathcal{T}^{\dagger} = \mbox{\boldmath $1$}_8$.  Now, define 
\begin{subequations}  
\label{transform1} 
\begin{equation} 
{\mathbb{F}} 
= {\mathcal{T}}{\mathcal{F}}  
= \frac{1}{2} 
\left(\begin{array}{c} 
\sqrt{\epsilon} E_x + \frac{i}{\sqrt{\mu}} B_x \\ 
\sqrt{\epsilon} E_y + \frac{i}{\sqrt{\mu}} B_y \\ 
\sqrt{\epsilon} E_z + \frac{i}{\sqrt{\mu}} B_z \\ 0 \\ 
i\left(\sqrt{\epsilon} E_x - \frac{i}{\sqrt{\mu}} B_x \right) \\ 
i\left(\sqrt{\epsilon} E_y - \frac{i}{\sqrt{\mu}} B_y \right) \\ 
i\left(\sqrt{\epsilon} E_z - \frac{i}{\sqrt{\mu}} B_z \right) \\ 0 
\end{array} \right) 
= \frac{1}{\sqrt{2}}
\left(\begin{array}{c}
F^{+}_x \\ F^{+}_y \\ F^{+}_z \\ 0 \\ iF^{-}_x \\ iF^{-}_y \\ iF^{-}_z \\ 0 \end{array}\right),  
\end{equation} 
\begin{equation} 
{\mathbb{J}} 
= {\mathcal{T}}{\mathcal{J}} 
= \frac{1}{2\sqrt{\epsilon}} 
\left(\begin{array}{c}
J_x \\  J_y \\  J_z \\ -iv \rho \\ iJ_x \\ iJ_y \\ iJ_z \\ -v \rho  
\end{array} \right).  
\end{equation} 
\end{subequations} 
Then, Eq. (\ref{dFdtMFJ-a}) becomes, in terms of ${\mathbb{F}}$ and ${\mathbb{J}}$,    
\begin{equation} 
\frac{\partial{\mathbb{F}}}{\partial t} 
= v\left({\mathcal{T}}{\mathcal{M}}_0{\mathcal{T}}^\dagger\right){\mathbb{F}} 
- {\mathbb{J}}      
= v\,\mathfrak{M}\,{\mathbb{F}} - {\mathbb{J}},  
\label{mathfrakMeqn} 
\end{equation} 
where 
\begin{eqnarray}       
\mathfrak{M} 
& = & {\mathcal{T}}{\mathcal{M}}_0{\mathcal{T}}^\dagger 
= \left(\begin{array}{cc}
\mathfrak{m} & \mbox{\boldmath $0$}_4 \\ 
\mbox{\boldmath $0$}_4 & \mathfrak{m}^* 
\end{array}\right),  \nonumber \\  
\mathfrak{m} 
& = & \left(\begin{array}{cccc} 
0 & i\partial_z & -i\partial_y & i\partial_x \\  
-i\partial_z &  0 & i\partial_x & i\partial_y \\
i\partial_y & -i\partial_x & 0 & i\partial_z \\             
-i\partial_x & -i\partial_y & -i\partial_z & 0      	       
\end{array} \right)  \nonumber \\ 
& = & \left(-\sigma_y \otimes \sigma_x\right)\partial_x 
+ \left(\sigma_y \otimes \sigma_z\right)\partial_y 
+ \left(-\mbox{\boldmath $1$}_2 \otimes \sigma_y\right)\partial_z,  
\label{mathfrakM}
\end{eqnarray} 
as follows from Eqs. (\ref{mathcalM0}), (\ref{mathcalM0xyz}), (\ref{tausigmatau}), and (\ref{Tau}).  Let 
\begin{equation}      
\mathfrak{m} = \mathfrak{m}_x\partial_x + \mathfrak{m}_y\partial_y 
+ \mathfrak{m}_z\partial_z.  
\end{equation}    
Then, it is seen that 
\begin{equation}
\mathfrak{m}_x \mathfrak{m}_y \mathfrak{m}_z  
= -i\mbox{\boldmath $1$}_4.  
\label{mxmymz}
\end{equation}   
Since $\sigma_x\sigma_y\sigma_z = i\mbox{\boldmath $1$}_2$, Eq. (\ref{mxmymz}) will be satisfied if $\mathfrak{m}_x$, $\mathfrak{m}_y$, and $\mathfrak{m}_z$ can be reduced to $-\mbox{\boldmath $1$}_2 \otimes \sigma_x$, 
$-\mbox{\boldmath $1$}_2 \otimes \sigma_y$, and 
$-\mbox{\boldmath $1$}_2 \otimes \sigma_x$, respectively, by a common similarity  transformation.  With this in mind, we shall seek a nonsingular matrix $U$ such that 
\begin{equation}
U \mathfrak{m}_x = -(\mbox{\boldmath $1$}_2 \otimes \sigma_x) U,  \quad 
U \mathfrak{m}_y = -(\mbox{\boldmath $1$}_2 \otimes \sigma_y) U,  \quad 
U \mathfrak{m}_z = -(\mbox{\boldmath $1$}_2 \otimes \sigma_z) U.  
\label{UmmU}
\end{equation} 
Since $\left\{\mathfrak{m}_x, \mathfrak{m}_y, \mathfrak{m}_z\right\}$ and  $\left\{-\mbox{\boldmath $1$}_2 \otimes \sigma_x, 
-\mbox{\boldmath $1$}_2 \otimes \sigma_y,  
-\mbox{\boldmath $1$}_2 \otimes \sigma_x\right\}$ are Hermitian we can expect $U$ 
to be unitary.  Substituting an arbitrary form of $U$,  
\begin{equation}
U = \left(\begin{array}{cccc} 
u_{11} & u_{12} & u_{13} & u_{14} \\ 
u_{21} & u_{22} & u_{23} & u_{24} \\ 
u_{31} & u_{32} & u_{33} & u_{34} \\ 
u_{41} & u_{42} & u_{43} & u_{44} 
\end{array}\right),  
\end{equation} 
in Eq. (\ref{UmmU}) it is possible to solve for the entries of $U$.  This process leads to a solution for all the entries of $U$ in terms of four undetermined entries $\left\{u_{11},u_{13},u_{31},u_{33}\right\}$.  We choose $u_{11} = -1/\sqrt{2}$, $u_{33} = 1/\sqrt{2}$ and $u_{13} = u_{31} = 0$ leading to the unitary matrix 
\begin{equation}
U = \frac{1}{\sqrt{2}}
\left(\begin{array}{cccc} 
-1 & i & 0 & 0 \\
0 & 0 & 1 & i \\ 
0 & 0 & 1 & -i \\ 
1 & i & 0 & 0 
\end{array}\right).  
\label{Uchoice}
\end{equation} 
Now, it can be verified directly that 
\begin{eqnarray}
U\mathfrak{m}U^{\dagger} 
& = & -\mbox{\boldmath $\Sigma$} \cdot \mbox{\boldmath $\nabla$} 
= -\left(\begin{array}{cc}
{\mbox{\boldmath $\sigma$}} \cdot {\mbox{\boldmath $\nabla$}} & 
{\mbox{\boldmath $0$}}_2 \\ 
{\mbox{\boldmath $0$}}_2 & {\mbox{\boldmath $\sigma$}} \cdot 
{\mbox{\boldmath $\nabla$}} 
\end{array} \right) 
= -{\mbox{\boldmath $1$}}_2 \otimes 
({\mbox{\boldmath $\sigma$}} \cdot {\mbox{\boldmath $\nabla$}}),  
\nonumber \\ 
{\mbox{\boldmath $\sigma$}} \cdot {\mbox{\boldmath $\nabla$}} 
& = & \sigma_x\partial_x + \sigma_y\partial_y + \sigma_z\partial_z 
\nonumber \\ 
& = & \left(\begin{array}{cc} 
\partial_z & \partial_x - i\partial_y \\ 
\partial_x + i\partial_y & -\partial_z 
\end{array}\right)  
= \left(\begin{array}{cc} 
\partial_z & \partial_{-} \\ 
\partial_{+} & -\partial_z 
\end{array}\right).         
\end{eqnarray} 
It is obvious that the lower diagonal block of $\mathfrak{M}$, namely $\mathfrak{m}^*$, will be reduced to a direct sum of two dimensional blocks, by the unitary matrix $U^*$.  However, we shall take the unitary matrix $-iU^*$ to block diagonalize $\mathfrak{m}^*$ such that 
\begin{eqnarray}
(-iU^*)\mathfrak{m}^*(iU^{*\,\dagger})  
& = & (U^*)\mathfrak{m}^*(U^{*\,\dagger})  
\nonumber \\ 
& = & -\mbox{\boldmath $\Sigma$}^* \cdot \mbox{\boldmath $\nabla$} 
= -\left(\begin{array}{cc}
{\mbox{\boldmath $\sigma$}}^* \cdot {\mbox{\boldmath $\nabla$}} & 
{\mbox{\boldmath $0$}}_2 \\ 
{\mbox{\boldmath $0$}}_2 & {\mbox{\boldmath $\sigma$}}^* \cdot 
{\mbox{\boldmath $\nabla$}} 
\end{array} \right)  \nonumber \\ 
& = & -{\mbox{\boldmath $1$}}_2 \otimes 
({\mbox{\boldmath $\sigma$}}^* \cdot {\mbox{\boldmath $\nabla$}}),  
\nonumber \\ 
{\mbox{\boldmath $\sigma$}}^* \cdot {\mbox{\boldmath $\nabla$}} 
& = & \left(\sigma_x\partial_x + \sigma_y\partial_y + \sigma_z\partial_z\right)^*  
\nonumber \\ 
& = & \left(\begin{array}{cc} 
\partial_z & \partial_x + i\partial_y \\ 
\partial_x - i\partial_y & -\partial_z 
\end{array}\right) 
= \left(\begin{array}{cc} 
\partial_z & \partial_{+} \\ 
\partial_{-} & -\partial_z 
\end{array}\right).     
\end{eqnarray}
Thus, with 
\begin{eqnarray}
T & = & \left(\begin{array}{cc}
U & \mbox{\boldmath $0$}_4 \\ \mbox{\boldmath $0$}_4 & -iU^* 
\end{array}\right)  
\nonumber \\  
& = & \frac{1}{\sqrt{2}}
\left(\begin{array}{rrrrrrrr} 
-1 & i & 0 &  0 &  0 &  0 &  0 &  0 \\
0 & 0 & 1 &  i &  0 &  0 &  0 &  0 \\ 
0 & 0 & 1 & -i &  0 &  0 &  0 &  0 \\
1 & i & 0 &  0 &  0 &  0 &  0 &  0 \\ 
0 & 0 & 0 &  0 &  i & -1 &  0 &  0 \\
0 & 0 & 0 &  0 &  0 &  0 & -i & -1 \\ 
0 & 0 & 0 &  0 &  0 &  0 & -i & 1 \\
0 & 0 & 0 &  0 & -i & -1 &  0 &  0 
\end{array} \right),  
\label{Tdefn} 
\end{eqnarray}  
we have 
\begin{equation}
T\mathfrak{M}T^{\dagger}  
= -\left(\begin{array}{cc} 
{\mbox{\boldmath $\Sigma$}}\cdot{\mbox{\boldmath $\nabla$}} & 
{\mbox{\boldmath $0$}} \\ 
{\mbox{\boldmath $0$}} & 
{\mbox{\boldmath $\Sigma^*$}}\cdot{\mbox{\boldmath$\nabla$}}    
\end{array}\right).   
\label{TmathfrakMT}
\end{equation} 
This completes the process of reduction of ${\mathcal{M}}_0$ in 
Eq. (\ref{dFdtMFJ-c}) to a direct sum of four irreducible two dimensional blocks.  The reason for the particular choice of the entries of $U$ and the presence of an extra $-i$ multiplying $U^*$ in the lower diagonal block of $T$ will be clear later. 

Let us now define 
\begin{equation} 
\Psi = T\,{\mathbb{F}} 
= \frac{1}{2} 
\left(\begin{array}{c} 
-F^{+}_x + iF^{+}_y \\ F^{+}_z \\ F^{+}_z \\ F^{+}_x + iF^{+}_y \\ 
-F_x^{-} - iF_y^{-} \\ F_z^{-} \\ F_z^{-} \\ F_x^{-} - iF_y^{-}  
\end{array}\right),  \qquad 
{\mathfrak{J}} = T{\mathbb{J}} 
= \frac{1}{2\sqrt{2 \epsilon}}  
\left(\begin{array}{c}
-J_x + iJ_y \\ J_z + v \rho \\ J_z - v \rho \\ J_x + iJ_y \\ 
-J_x - iJ_y \\ J_z + v \rho \\ J_z - v \rho \\ J_x - iJ_y  
\end{array} \right).   
\label{transform2} 
\end{equation}  
Then, from Eqs. (\ref{mathfrakMeqn}) and (\ref{TmathfrakMT}) it follows that   
\begin{eqnarray} 	 
\frac{\partial \Psi}{\partial t} 
& = & v\left(T\,\mathfrak{M}\,T^\dagger\right)\Psi - {\mathfrak{J}}  
= vM_0\,\Psi - {\mathfrak{J}},  
\nonumber \\ 
M_0 
& = & -\left(\begin{array}{cc} 
{\mbox{\boldmath $\Sigma$}}\cdot{\mbox{\boldmath $\nabla$}} & 
{\mbox{\boldmath $0$}} \\ 
{\mbox{\boldmath $0$}} & 
{\mbox{\boldmath $\Sigma^*$}}\cdot{\mbox{\boldmath$\nabla$}}    
\end{array}\right) 
\nonumber \\ 
& = & -\left(\begin{array}{cccccccc} 
\partial_z & \partial_{-} & 0 & 0 & 0 & 0 & 0 & 0 \\ 
\partial_{+} & -\partial_z & 0 & 0 & 0 & 0 & 0 & 0 \\ 
0 & 0 & \partial_z & \partial_{-} & 0 & 0 & 0 & 0 \\ 
0 & 0 &  \partial_{+} & -\partial_z & 0 & 0 & 0 & 0 \\ 
0 & 0 & 0 & 0 & \partial_z & \partial_{+} & 0 & 0 \\ 
0 & 0 & 0 & 0 & \partial_{-} & -\partial_z & 0 & 0 \\ 
0 & 0 & 0 & 0 & 0 & 0 & \partial_z & \partial_{+} \\ 
0 & 0 & 0 & 0 & 0 & 0 & \partial_{-} & -\partial_z 
\end{array} \right).       
\label{PsiEvolnEq}
\end{eqnarray}  
The vectors  
\begin{equation} 
{\mbox{\boldmath $F$}}^{\pm}   
= \frac{1}{\sqrt{2}}\left(\sqrt{\epsilon}{\mbox{\boldmath $E$}} 
\pm \frac{i}{\sqrt{\mu}}\mbox{\boldmath $B$}\right) 
= \sqrt{\frac{\epsilon}{2}}({\mbox{\boldmath $E$}} 
\pm iv{\mbox{\boldmath $B$}}),  
\end{equation}  
appearing in the definition of $\mathbb{F}$ in Eqs. (\ref{transform1}) and 
(\ref{transform2}), are the Riemann-Silberstein-Weber (RSW) vectors.  For details on the properties of the RSW vectors, including their transformation properties under rotations and Lorentz transformations, see, {\em e.g.}, (Bialynicki-Birula, 1994, 1996a,b;  Bialynicki-Birula and Bialynicki-Birula, 2013; Przeszowski, 2023).  Expressed in terms of $n$ and $\eta$, or $v$ and $\eta$, 
\begin{equation} 
{\mbox{\boldmath $F$}}^{\pm} 
= \frac{1}{\sqrt{2}}\left(\sqrt{\frac{n}{c\eta}}\,{\mbox{\boldmath $E$}} 
\pm i\sqrt{\frac{c}{n\eta}}\,{\mbox{\boldmath $B$}}\right)  
= \frac{1}{\sqrt{2v\eta}}\left({\mbox{\boldmath $E$}} 
\pm iv {\mbox{\boldmath $B$}}\right).       
\end{equation}     

It should be pointed out that in the process of complete reduction of $\mathcal{M}_0$ we have obtained the RSW vectors, ${\mbox{\boldmath $F$}^{+}}$ and ${\mbox{\boldmath $F$}^{-}}$, as the elements of $\Psi$, the basis vector of the representation.  It may be noted that ${\mbox{\boldmath $F$}^{+}}$ and ${\mbox{\boldmath $F$}^{-}}$ are linear combinations of ${\mbox{\boldmath $E$}}$ and ${\mbox{\boldmath $B$}}$ with $\pm i$ as coefficients of 
${\mbox{\boldmath $B$}}/\sqrt{\mu}$ irrespective of whether the fields are written in real notation or complex notation.  The superscripts $+$ and $-$ in 
${\mbox{\boldmath $F$}^{+}}$ and ${\mbox{\boldmath $F$}^{-}}$, respectively, refer to the signs $\pm$ of the coefficient for ${\mbox{\boldmath $B$}}/\sqrt{\mu}$.  Consequently, ${\mbox{\boldmath $F$}^{+}}$ and ${\mbox{\boldmath $F$}^{-}}$ are not mutually complex conjugate.  For the same reason, the lower four components of $\Psi$ are not the complex conjugate counterparts of the upper four components.    

To proceed further, it is instructive to introduce 
\begin{subequations}  
\label{psi+-j+-defn} 
\begin{equation} 
\Psi 
= \left(\begin{array}{c} 
\psi^{+} \\ \psi^{-}
\end{array}\right),  \qquad 
{\mathfrak{J}} 
= \left(\begin{array}{l} 
{\mathfrak{j}}^{+} \\ {\mathfrak{j}}^{-}
\end{array} \right),  
\end{equation} 
\begin{equation} 
\psi^{+} 
= \frac{1}{2}\left(\begin{array}{c} 
-F^{+}_x + iF^{+}_y \\ F^{+}_z \\ F^{+}_z \\ F^{+}_x + iF^{+}_y,    
\end{array} \right), \qquad 
\psi^{-} 
= \frac{1}{2}\left(\begin{array}{c} 
-F^{-}_x - iF^{-}_y \\ F^{-}_z \\ F^{-}_z \\ F^{-}_x - iF^{-}_y, 
\end{array} \right),  
\end{equation}  
\begin{equation} 
{\mathfrak{j}}^{+} 
= \frac{1}{2\sqrt{2\epsilon}}  
\left(\begin{array}{c}
-J_x + iJ_y \\ J_z + v\rho \\ J_z - v\rho \\ J_x + iJ_y  
\end{array} \right),  \qquad 
{\mathfrak{j}}^{-} 
= \frac{1}{2\sqrt{2\epsilon}}  
\left(\begin{array}{c} 
-J_x - iJ_y \\ J_z + v\rho \\ J_z - v\rho \\ J_x - iJ_y  
\end{array} \right).
\end{equation}  	              
\end{subequations}  
Then, we can write Eq. (\ref{PsiEvolnEq}) as a pair of equations 
\begin{equation} 
\frac{\partial\psi^{+}}{\partial t} 
= -v({\mbox{\boldmath $\Sigma$}}\cdot
{\mbox{\boldmath $\nabla$}})\psi^{+} - {\mathfrak{j}}^{+},  \qquad  
\frac{\partial\psi^{-}}{\partial t} 
= -v({\mbox{\boldmath $\Sigma$}^*}\cdot
{\mbox{\boldmath $\nabla$}})\psi^{-} - {\mathfrak{j}}^{-}.  
\label{psieqns4vcum} 
\end{equation} 
Substitution of the expressions for $\psi^{+}$, $\mathfrak{j}^{+}$, $\psi^{-}$, and $\mathfrak{j}^{-}$, in Eq. (\ref{psieqns4vcum}) leads to 
\begin{subequations} 
\begin{equation} 
\frac{\partial {\mbox{\boldmath $F$}}^{+}}{\partial t} 
= -iv({\mbox{\boldmath $\nabla$}}\times{\mbox{\boldmath $F$}}^{+}) 
-\frac{1}{\sqrt{2\epsilon}}{\mbox{\boldmath $J$}},  \qquad  
{\mbox{\boldmath $\nabla$}}\cdot{\mbox{\boldmath $F$}}^{+} 
= \frac{\rho}{\sqrt{2 \epsilon}},       
\label{RSWMEq+} 
\end{equation}  
\begin{equation} 
\frac{\partial {\mbox{\boldmath $F$}}^{-}}{\partial t} 
= iv({\mbox{\boldmath $\nabla$}}\times{\mbox{\boldmath $F$}}^{-}) -\frac{1}{\sqrt{2\epsilon}}{\mbox{\boldmath $J$}},  \qquad 
{\mbox{\boldmath $\nabla$}} \cdot {\mbox{\boldmath $F$}}^{-}  
= \frac{\rho}{\sqrt{2 \epsilon}}.  
\label{RSWMEq-} 
\end{equation} 
\end{subequations} 
Equations (\ref{RSWMEq+}) and (\ref{RSWMEq-}) represent the Maxwell equations for a homogeneous medium in terms of the RSW-vectors.  In the absence of charges and currents we have    
\begin{equation}
\frac{\partial{\mbox{\boldmath $F$}}^{\pm}}{\partial t} 
= \mp iv({\mbox{\boldmath $\nabla$}}\times{\mbox{\boldmath $F$}}^{\pm}),  \qquad 
{\mbox{\boldmath $\nabla$}}\cdot{\mbox{\boldmath $F$}}^{\pm} = 0,  
\label{SfreeRSWMEq} 
\end{equation} 

From Eqs. (\ref{transform1}) and (\ref{transform2}) we have 
\begin{eqnarray}
\Psi & = & {\mathbb{T}}{\mathcal{F}}, \qquad 
{\mathfrak{J}} = {\mathbb{T}}{\mathcal{J}},  
\nonumber \\ 
{\mathbb{T}} 
& = & T{\mathcal{T}} 
= \frac{1}{2}
\left(\begin{array}{rrrrrrrr}
-1 &  i & 0 &  0 & -i & -1 &  0 &  0 \\
0 &  0 & 1 &  i &  0 &  0 &  i & -1 \\ 
0 &  0 & 1 & -i &  0 &  0 &  i &  1 \\
1 &  i & 0 &  0 &  i & -1 &  0 &  0 \\ 
-1 & -i & 0 &  0 &  i & -1 &  0 &  0 \\
0 &  0 & 1 & -i &  0 &  0 & -i & -1 \\ 
0 &  0 & 1 &  i &  0 &  0 & -i &  1 \\
1 & -i & 0 &  0 & -i & -1 &  0 &  0 
\end{array} \right),     
\label{F2PsiJ2J} 
\end{eqnarray}  
where ${\mathbb{T}}$ is unitary.  Thus, we can retrieve $\mathcal{F}$ and $\mathcal{J}$, or in other words  ${\mbox{\boldmath $E$}}$, 
${\mbox{\boldmath $B$}}$, ${\mbox{\boldmath $J$}}$, and $\rho$, from $\Psi$ and $\mathfrak{J}$ by the inverse transformations  
\begin{equation} 
{\mathcal{F}} = {\mathbb{T}}^\dagger\Psi, \qquad 
{\mathcal{J}} = {\mathbb{T}}^\dagger{\mathfrak{J}}.  
\label{InvTransf}
\end{equation}       

Quantum mechanical description of the photon, associating the photon with a wavefunction, has been discussed by several authors based on the RSW vector (see, {\em e.g.},   Bialynicki-Birula, 1994, 1996a,b; Bialynicki-Birula and Bialynicki-Birula, 2013; Barnett, 2014; Kiesslinga and Tahvildar-Zadehb, 2018; Sebens, 2019, for a history of the much debated topic and references).  In this connection the following may be noted.  Both $\sqrt{\epsilon}{\mbox{\boldmath $E$}}$ and ${\mbox{\boldmath$B$}}/\sqrt{\mu}$ in ${\mathcal{F}}$ have the dimension of square root of energy, with 
\begin{equation} 
{\mathcal{F}}^{\dagger}({\mbox{\boldmath $r$}},t)  
{\mathcal{F}}({\mbox{\boldmath $r$}},t) 
= \frac{1}{2}\left(\epsilon|{\mbox{\boldmath $E$}}({\mbox{\boldmath $r$}},t)|^2 
+\frac{1}{\mu}|{\mbox{\boldmath $B$}}({\mbox{\boldmath $r$}},t)|^2\right),   
\end{equation} 
being the local energy density at the position $\mbox{\boldmath $r$}$ and at time $t$, resembling the quantum mechanical position probability density.  It is this resemblance that leads to the suggestion of quantum mechanical theory of the photon.  If we let 
\begin{equation} 
{\mathcal{F}}({\mbox{\boldmath $r$}},t) 
= \langle{\mbox{\boldmath $r$}}|{\mathcal{F}}(t)\rangle,  
\end{equation} 
then we have 
\begin{eqnarray} 
\langle {\mathcal{F}}(t)|{\mathcal{F}}(t)\rangle 
& = & \int_{-\infty}^{\infty}\int_{-\infty}^{\infty}\int_{-\infty}^{\infty} 
dx\,dy\,dz\,{\mathcal{F}}({\mbox{\boldmath $r$}},t)^\dagger {\mathcal{F}}({\mbox{\boldmath $r$}},t)  
\nonumber \\ 
& = & \frac{1}{2}\int_{-\infty}^{\infty}\int_{-\infty}^{\infty}
\int_{-\infty}^{\infty}dx\,dy\,dz\,\left(\epsilon 
|{\mbox{\boldmath $E({\mbox{\boldmath $r$}},t)$}}|^2 
+ \frac{1}{\mu}|{\mbox{\boldmath $B({\mbox{\boldmath $r$}},t)$}}|^2\right),  \nonumber \\ 
   &   &  
\end{eqnarray} 
the total energy of the field at time $t$.  Write the time evolution equation for $\mathcal{F}(t)$ in a time-independent source-free homogeneous medium,  
Eq. (\ref{dFdtMFJ-a}) with $\mathcal{J} = 0$, as 
\begin{equation} 
\frac{\partial|{\mathcal{F}}(t)\rangle}{\partial t} 
= \mathcal{M}_0|{\mathcal{F}(t)}\rangle.      
\end{equation} 
Then, since $\mathcal{M}_0$ is time-independent,   
\begin{equation} 
|{\mathcal{F}(t)} \rangle 
= \exp{\left(\mathcal{M}_0 t\right)}|{\mathcal{F}(0)} \rangle, \qquad 
\langle {\mathcal{F}(t)}| 
= \langle {\mathcal{F}(0)}
|\left(\exp{\left(\mathcal{M}_0 t \right)}\right)^\dagger.  
\end{equation} 
Since ${\mathcal{M}}_0$ is antihermitian, as already established, 
$\left(\exp{\left(\mathcal{M}_0 t\right)}\right)^\dagger 
= \exp{\left(-\mathcal{M}_0 t\right)}$.  Consequently, 
$\exp{\left(\mathcal{M}_0 t\right)}$ is unitary, and   
\begin{equation} 
\langle {\mathcal{F}}(t)|{\mathcal{F}}(t) \rangle 
= \langle {\mathcal{F}}(0)|{\mathcal{F}}(0) \rangle. 
\end{equation} 
Thus, the time evolution of ${\mathcal{F}}(t)$ is unitary and the total energy of the field is conserved in a time-independent source-free homogeneous medium.  Since $\Psi = \mathbb{T}\mathcal{F}$, with unitary $\mathbb{T}$, $\Psi$ has also unitary time evolution in a time-independent source-free homogeneous medium as seen in Eq. (\ref{PsiEvolnEq}).     

\section{The new matrix representation of the Maxwell equations for a linear inhomogeneous medium}  
 
If $\epsilon$ and $\mu$ are varying with space and time, then we can write  
Eq. (\ref{MEqepsmu}) as  
\begin{eqnarray}
&   & \frac{\partial(\sqrt{\epsilon}{\mbox{\boldmath $E$}})}{\partial t} 
= -\frac{\dot{\epsilon}}{2\epsilon}(\sqrt{\epsilon}
{\mbox{\boldmath $E$}}) + v\left({\mbox{\boldmath $\nabla$}} -\frac{1}{2\mu}{\mbox{\boldmath $\nabla$}}\mu\right) 
\times\left(\frac{{\mbox{\boldmath $B$}}}{\sqrt{\mu}}\right) 
-\frac{{\mbox{\boldmath $J$}}}{\sqrt{\epsilon}},  
\nonumber \\  
&   & \left({\mbox{\boldmath $\nabla$}} 
+ \frac{1}{2\epsilon}{\mbox{\boldmath $\nabla$}}\epsilon\right)\cdot
\left(\sqrt{\epsilon}{\mbox{\boldmath $E$}}\right) 
= \frac{\rho}{\sqrt{\epsilon}},  
\nonumber \\ 
&   & \frac{\partial}{\partial t}
\left(\frac{{\mbox{\boldmath $B$}}}{\sqrt{\mu}}\right) 
= -\frac{\dot{\mu}}{2\mu}\left(\frac{{\mbox{\boldmath $B$}}}{\sqrt{\mu}}\right) 
- v\left({\mbox{\boldmath $\nabla$}} 
-\frac{1}{2\epsilon}{\mbox{\boldmath $\nabla$}}\epsilon\right)\times
\left(\sqrt{\epsilon}{\mbox{\boldmath $E$}}\right),  
\nonumber \\   
&   & \left({\mbox{\boldmath $\nabla$}} 
+ \frac{1}{2\mu}{\mbox{\boldmath $\nabla$}}\mu\right)     
\cdot\left(\frac{{\mbox{\boldmath $B$}}}{\sqrt{\mu}}\right) 
= 0,  
\label{NCMEqepsmu} 
\end{eqnarray}
instead of Eq. (\ref{MEqconstepsmu}).  Defining  $\bar{\epsilon} = \ln{\epsilon}/2$ and $\bar{\mu} = \ln{\mu}/2$, Eq. (\ref{NCMEqepsmu}) can be expressed in terms of $\mathcal{F}$ and $\mathcal{J}$ as   
\begin{subequations} 
\label{NCdFdt} 
\begin{equation} 
\frac{\partial\mathcal{F}}{\partial t} 
= \mathcal{M}{\mathcal{F}} - {\mathcal{J}},  \quad  
\mathcal{M} 
= \left(\begin{array}{cc}
{\mathcal{M}}_{11} & {\mathcal{M}}_{12} \\ 
{\mathcal{M}}_{21} & {\mathcal{M}}_{22} 
\end{array} \right),  
\end{equation}
\begin{equation}
{\mathcal{M}}_{11} = -\dot{\bar{\epsilon}}{\mbox{\boldmath $1$}}_4,  \quad 
{\mathcal{M}}_{22} = -\dot{\bar{\mu}}{\mbox{\boldmath $1$}}_4,   	
\end{equation}
\begin{equation}
{\mathcal{M}}_{12} 
= v\left(\begin{array}{cccc} 
0 & -\partial_z + \partial_z\bar{\mu} 
& \partial_y - \partial_y\bar{\mu} & -\partial_x - \partial_x\bar{\mu} \\ 
\partial_z - \partial_z\bar{\mu} & 0 
& -\partial_x + \partial_x\bar{\mu} & -\partial_y - \partial_y\bar{\mu} \\ 
-\partial_y + \partial_y\bar{\mu} &  \partial_x - \partial_x\bar{\mu} 
& 0 & -\partial_z - \partial_z\bar{\mu} \\ 
\partial_x + \partial_x\bar{\mu} &  \partial_y + \partial_y\bar{\mu} 
& \partial_z + \partial_z\bar{\mu} & 0 
\end{array} \right),  
\end{equation}
\begin{equation}
{\mathcal{M}}_{21} 
= v\left(\begin{array}{cccc}
0 & \partial_z - \partial_z\bar{\epsilon}  
& -\partial_y + \partial_y\bar{\epsilon} & \partial_x + \partial_x\bar{\epsilon} \\ 
-\partial_z + \partial_z\bar{\epsilon} & 0 
& \partial_x - \partial_x\bar{\epsilon} & \partial_y + \partial_y\bar{\epsilon} \\ 
\partial_y - \partial_y\bar{\epsilon} & -\partial_x + \partial_x\bar{\epsilon}
& 0 & \partial_z + \partial_z\bar{\epsilon} \\ 
-\partial_x - \partial_x\bar{\epsilon} & -\partial_y - \partial_y\bar{\epsilon} 
& -\partial_z - \partial_z\bar{\epsilon} & 0 
\end{array} \right),    
\end{equation}  	
\end{subequations}   
where $v$ varies with space and time.  When $\epsilon$, $\mu$, and hence $v$, are time-independent and the medium is source-free we have 
\begin{equation}
\frac{\partial\mathcal{F}}{\partial t} 
= \mathcal{M}{\mathcal{F}}   
= \left(\begin{array}{cc}
{\mbox{\boldmath $0$}} & {\mathcal{M}}_{12} \\ 
{\mathcal{M}}_{21} & {\mbox{\boldmath $0$}}   
\end{array}\right){\mathcal{F}}.       
\label{tiNCdFdt} 
\end{equation}  
For such a medium with time-independent $\mathcal{M}$ we can write 
\begin{eqnarray} 
\frac{\partial^2{\mathcal{F}}}{\partial t^2} 
& = & \left(\begin{array}{cc}
{\mbox{\boldmath $0$}} & {\mathcal{M}}_{12} \\ 
{\mathcal{M}}_{21} & {\mbox{\boldmath $0$}}  
\end{array}\right)
\left(\begin{array}{cc}
{\mbox{\boldmath $0$}} & {\mathcal{M}}_{12} \\ 
{\mathcal{M}}_{21} & {\mbox{\boldmath $0$}}   
\end{array} \right){\mathcal{F}}  
\nonumber \\  
& = & \left(\begin{array}{cc}
{\mathcal{M}}_{12}{\mathcal{M}}_{21} & {\mbox{\boldmath $0$}} \\ 
{\mbox{\boldmath $0$}} & {\mathcal{M}}_{21}{\mathcal{M}}_{12}   
\end{array} \right){\mathcal{F}}.     
\label{tiNCd2Fdt2}
\end{eqnarray}   
As shown in Appendix B this equation leads to the generalized Helmholtz equations for ${\mbox{\boldmath $E$}}$ and ${\mbox{\boldmath $B$}}$ given in 
(Born and Wolf, 1999; Mazharimousavi et al., 2013)  which reduce to the Helmholtz equations in Eq. (\ref{Helmholtz}) for the case when $\epsilon$ and $\mu$ are constants.  It may be noted that when $\epsilon$ and $\mu$ are constants 
${\mathcal{M}} \longrightarrow {\mathcal{M}}_0$.  

Writing Eq. (\ref{NCdFdt}) in terms of $\Psi$ and $\mathfrak{J}$, using 
Eq. (\ref{F2PsiJ2J}), we get 
\begin{subequations} 
\label{NCdPsidtepsmu} 
\begin{equation}
\frac{\partial \Psi}{\partial t} 
= \left(\mathbb{T}\mathcal{M}\mathbb{T}^\dagger\right){\Psi} - {\mathfrak{J}} 
= M{\Psi} - {\mathfrak{J}}, \quad  
M = vM_0 + M^\prime,  
\end{equation}
\begin{equation} 
M_0 = -\left(\begin{array}{cc} 
{\mbox{\boldmath $\Sigma$}}\cdot{\mbox{\boldmath $\nabla$}} & 
{\mbox{\boldmath $0$}} \\ 
{\mbox{\boldmath $0$}} & 
{\mbox{\boldmath $\Sigma$}}^*\cdot{\mbox{\boldmath $\nabla$}}  
\end{array} \right),   
\end{equation}	
\begin{equation}  
M^\prime 
= \frac{v}{2}\left(\begin{array}{cc} 
\begin{array}{l} 
({\mbox{\boldmath $\sigma$}}\cdot
({\mbox{\boldmath $\nabla$}}(\bar{\epsilon}+\bar{\mu}))
\otimes{\mbox{\boldmath $1$}}_2) \\  
-\left(\frac{\dot{\epsilon} + \dot{\mu}}{v}\right){\mbox{\boldmath $1$}}_4  
\end{array} & 
\begin{array}{l} 
(({\mbox{\boldmath $\sigma$}}\cdot({\mbox{\boldmath $\nabla$}}
(\bar{\epsilon}-\bar{\mu}))\sigma_y)\otimes\sigma_y) \\ 
-\left(\frac{\dot{\epsilon}-\dot{\mu}}{v}\right)
\left(\sigma_y\otimes\sigma_y\right) 
\end{array} \\ & \\
\begin{array}{l} 
(({\mbox{\boldmath $\sigma$}}^*\cdot({\mbox{\boldmath $\nabla$}}
(\bar{\epsilon}-\bar{\mu}))\sigma_y)\otimes\sigma_y) \\ 
-\left(\frac{\dot{\epsilon}-\dot{\mu}}{v}\right)
\left(\sigma_y\otimes\sigma_y\right)    
\end{array} &  
\begin{array}{l} 
(({\mbox{\boldmath $\sigma$}}^*\cdot
({\mbox{\boldmath $\nabla$}}(\bar{\epsilon}+\bar{\mu})) 
\otimes{\mbox{\boldmath $1$}}_2)) \\  
-\left(\frac{\dot{\epsilon} + \dot\mu}{v}\right){\mbox{\boldmath $1$}}_4   
\end{array}  
\end{array} \right).  
\end{equation} 
\end{subequations} 
Explicitly writing, we have   
\begin{eqnarray} 
M^\prime 
& = & \left(\begin{array}{cc} 
M^\prime_{11} & M^\prime_{12} \\ 
M^\prime_{21} & M^\prime_{22} 
\end{array}\right),  
\nonumber \\ 
M^\prime_{11} 
& = & \frac{v}{2}\left( 
\begin{array}{cccc} 
\begin{array}{c}
\partial_z(\bar{\epsilon}+\bar{\mu}) \\ 
-\frac{1}{v}(\dot{\bar{\epsilon}}+\dot{\bar{\mu}}) 
\end{array} & 0 & \partial_{-}(\bar{\epsilon}+\bar{\mu}) & 0 \\ 
0 & \begin{array}{c}
\partial_z(\bar{\epsilon}+\bar{\mu}) \\ 
-\frac{1}{v}(\dot{\bar{\epsilon}}+\dot{\bar{\mu}}) 
\end{array} & 0 & \partial_{-}(\bar{\epsilon}+\bar{\mu}) \\ 
\partial_{+}(\bar{\epsilon}+\bar{\mu}) & 0 & 
\begin{array}{c}
-\partial_z(\bar{\epsilon}+\bar{\mu}) \\ 
-\frac{1}{v}(\dot{\bar{\epsilon}}+\dot{\bar{\mu}}) 
\end{array} & 0 \\ 
0 & \partial_{+}(\bar{\epsilon}+\bar{\mu}) & 0 & 
\begin{array}{c}
-\partial_z(\bar{\epsilon}+\bar{\mu}) \\ 
-\frac{1}{v}(\dot{\bar{\epsilon}}+\dot{\bar{\mu}}) 
\end{array} 
\end{array}\right),  
\nonumber \\   
M^\prime_{12} 
& = & \frac{v}{2}\left( 
\begin{array}{cccc} 
0 & \partial_{-}(\bar{\epsilon}-\bar{\mu}) & 0 & 
\begin{array}{c} 
-\partial_z(\bar{\epsilon}-\bar{\mu}) \\
+\frac{1}{v}(\dot{\bar{\epsilon}}-\dot{\bar{\mu}}) 
\end{array} \\ 
-\partial_{-}(\bar{\epsilon}-\bar{\mu}) & 0 & 
\begin{array}{c}
\partial_z(\bar{\epsilon}-\bar{\mu}) \\ 
-\frac{1}{v}(\dot{\bar{\epsilon}}-\dot{\bar{\mu}}) 
\end{array} & 0 \\ 
0 & \begin{array}{c}
-\partial_z(\bar{\epsilon}-\bar{\mu}) \\ 
-\frac{1}{v}(\dot{\bar{\epsilon}}-\dot{\bar{\mu}}) 
\end{array} & 0 & -\partial_{+}(\bar{\epsilon}-\bar{\mu}) \\ 
\begin{array}{c}
\partial_z(\bar{\epsilon}-\bar{\mu}) \\ 
+\frac{1}{v}(\dot{\bar{\epsilon}}-\dot{\bar{\mu}}) 
\end{array} & 0 & \partial_{+}(\bar{\epsilon}-\bar{\mu}) & 0 
\end{array}\right),  
\nonumber \\ 
M^\prime_{21} 
& = & \frac{v}{2}\left( 
\begin{array}{cccc} 
0 & \partial_{+}(\bar{\epsilon}-\bar{\mu}) & 0 & 
\begin{array}{c} 
-\partial_z(\bar{\epsilon}-\bar{\mu}) \\
+\frac{1}{v}(\dot{\bar{\epsilon}}-\dot{\bar{\mu}}) 
\end{array} \\ 
-\partial_{+}(\bar{\epsilon}-\bar{\mu}) & 0 & 
\begin{array}{c}
\partial_z(\bar{\epsilon}-\bar{\mu}) \\ 
-\frac{1}{v}(\dot{\bar{\epsilon}}-\dot{\bar{\mu}}) 
\end{array} & 0 \\ 
0 & \begin{array}{c}
-\partial_z(\bar{\epsilon}-\bar{\mu}) \\ 
-\frac{1}{v}(\dot{\bar{\epsilon}}-\dot{\bar{\mu}}) 
\end{array} & 0 & -\partial_{-}(\bar{\epsilon}-\bar{\mu}) \\ 
\begin{array}{c}
\partial_z(\bar{\epsilon}-\bar{\mu}) \\ 
+\frac{1}{v}(\dot{\bar{\epsilon}}-\dot{\bar{\mu}}) 
\end{array} & 0 & \partial_{-}(\bar{\epsilon}-\bar{\mu}) & 0 
\end{array}\right),  
\nonumber \\  
M^\prime_{22} 
& = & \frac{v}{2}\left( 
\begin{array}{cccc} 
\begin{array}{c}
\partial_z(\bar{\epsilon}+\bar{\mu}) \\ 
-\frac{1}{v}(\dot{\bar{\epsilon}}+\dot{\bar{\mu}}) 
\end{array} & 0 & \partial_{+}(\bar{\epsilon}+\bar{\mu}) & 0 \\ 
0 & \begin{array}{c}
\partial_z(\bar{\epsilon}+\bar{\mu}) \\ 
-\frac{1}{v}(\dot{\bar{\epsilon}}+\dot{\bar{\mu}}) 
\end{array} & 0 & \partial_{+}(\bar{\epsilon}+\bar{\mu}) \\ 
\partial_{-}(\bar{\epsilon}+\bar{\mu}) & 0 & 
\begin{array}{c}
-\partial_z(\bar{\epsilon}+\bar{\mu}) \\ 
-\frac{1}{v}(\dot{\bar{\epsilon}}+\dot{\bar{\mu}}) 
\end{array} & 0 \\ 
0 & \partial_{-}(\bar{\epsilon}+\bar{\mu}) & 0 & 
\begin{array}{c}
-\partial_z(\bar{\epsilon}+\bar{\mu}) \\ 
-\frac{1}{v}(\dot{\bar{\epsilon}}+\dot{\bar{\mu}}) 
\end{array} 
\end{array}\right).  
\nonumber \\ 
&   &  
\end{eqnarray} 

Let us now write Eq. (\ref{NCdPsidtepsmu}) in terms of the refractive index $n$ and the impedance $\eta$.  To this end, we define $\bar{n} = \ln{n}/2$ and 
$\bar{\eta} = \ln{\eta}/2$ and use the relations 
\begin{equation} 
\epsilon = \frac{n}{c\eta}, \quad 
\mu = \frac{n\eta}{c}, 
\quad  v = \frac{c}{n}.  
\end{equation} 
Then, we have 
\begin{eqnarray}
\dot{\bar{\epsilon}}+\dot{\bar{\mu}} 
& = & 2\dot{\bar{n}},  \quad  
\dot{\bar{\epsilon}}-\dot{\bar{\mu}} 
= -2\dot{\bar{\eta}},  \nonumber \\ 
{\mbox{\boldmath $\nabla$}}(\bar{\epsilon}+\bar{\mu}) 
& = & 2{\mbox{\boldmath $\nabla$}}\bar{n},  \quad 
{\mbox{\boldmath $\nabla$}}(\bar{\epsilon}-\bar{\mu}) 
= -2{\mbox{\boldmath $\nabla$}}\bar{\eta}.  
\end{eqnarray} 
Using these relations, Eq. (\ref{NCdPsidtepsmu}) becomes    
\begin{subequations} 
\label{theEqn}  
\begin{equation}
\frac{\partial \Psi}{\partial t} = M\Psi - {\mathfrak{J}}, \quad  
M = \frac{c}{n}\,M_0 + M^\prime,  
\end{equation} 
\begin{equation}
M^\prime 
= \frac{c}{n}\left( \begin{array}{cc}  
(({\mbox{\boldmath $\sigma$}}\cdot({\mbox{\boldmath $\nabla$}}\bar{n}))
\otimes{\mbox{\boldmath $1$}}_2) 
-\left(\frac{\dot{n}}{2c}\right){\mbox{\boldmath $1$}}_4 &  
\begin{array}{l} 
-\left(\left(({\mbox{\boldmath $\sigma$}}\cdot
({\mbox{\boldmath $\nabla$}}\bar{\eta}))\sigma_y\right)
\otimes\sigma_y\right) \\ 
+\frac{n\dot{\eta}}{2c\eta}\left(\sigma_y\otimes\sigma_y\right)    
\end{array} \\ & \\ 
\begin{array}{l} 
-\left(\left(\left({\mbox{\boldmath $\sigma$}}^*\cdot
({\mbox{\boldmath $\nabla$}}\bar{\eta})\right)\sigma_y\right)
\otimes\sigma_y\right) \\ 
+\frac{n\dot{\eta}}{2c\eta}\left(\sigma_y\otimes\sigma_y\right)   
\end{array} & 
(({\mbox{\boldmath $\sigma$}}^*\cdot
({\mbox{\boldmath $\nabla$}}\bar{n}))\otimes{\mbox{\boldmath $1$}}_2)  
-\left(\frac{\dot{n}}{2c}\right){\mbox{\boldmath $1$}}_4    
\end{array} \right).  	
\end{equation}
\end{subequations} 
Explicitly writing, we have    
\begin{eqnarray}   
M^\prime_{11} 
& = & \frac{c}{n}\left( 
\begin{array}{cccc} 
\partial_z\bar{n}-\frac{\dot{n}}{2c} & 0 & \partial_{-}\bar{n} & 0 \\ 
0 & \partial_z\bar{n}-\frac{\dot{n}}{2c} & 0 & \partial_{-}\bar{n} \\ 
\partial_+\bar{n} & 0 & -\partial_z\bar{n}-\frac{\dot{n}}{2c} & 0 \\ 
0 & \partial_+\bar{n} & 0 & -\partial_z\bar{n}-\frac{\dot{n}}{2c}  
\end{array} \right),  
\nonumber \\ 
M^\prime_{12} 
& = & \frac{c}{n}\left( 
\begin{array}{cccc} 
0 & -\partial_{-}\bar{\eta} & 0 & \partial_z\bar{\eta}-\frac{n\dot{\eta}}{2c\eta} \\ 
\partial_{-}\bar{\eta} & 0 & -\partial_z\bar{\eta}+\frac{n\dot{\eta}}{2c\eta} & 0 \\ 
0 & \partial_z\bar{\eta}+\frac{n\dot{\eta}}{2c\eta} & 0 & \partial_+\bar{\eta} \\ 
-\partial_z\bar{\eta}-\frac{n\dot{\eta}}{2c\eta} & 0 & -\partial_+\bar{\eta} & 0  
\end{array} \right),   
\nonumber \\ 
M^\prime_{21} 
& = & \frac{c}{n}\left( 
\begin{array}{cccc} 
0 & -\partial_+\bar{\eta} & 0 & \partial_z\bar{\eta}-\frac{n\dot{\eta}}{2c\eta} \\ 
\partial_+\bar{\eta} & 0 & -\partial_z\bar{\eta}+\frac{n\dot{\eta}}{2c\eta} & 0 \\ 
0 & \partial_z\bar{\eta}+\frac{n\dot{\eta}}{2c\eta} & 0 & \partial_{-}\bar{\eta} \\ 
-\partial_z\bar{\eta}-\frac{n\dot{\eta}}{2c\eta} & 0 & -\partial_{-}\bar{\eta} & 0   
\end{array} \right),  
\nonumber \\  
M^\prime_{22} 
& = & \frac{c}{n}\left( 
\begin{array}{cccc} 
\partial_z\bar{n}-\frac{\dot{n}}{2c} & 0 & \partial_{+}\bar{n} & 0 \\ 
0 & \partial_z\bar{n}-\frac{\dot{n}}{2c} & 0 & \partial_{+}\bar{n} \\ 
\partial_{-}\bar{n} & 0 & -\partial_z\bar{n}-\frac{\dot{n}}{2c} & 0 \\ 
0 & \partial_{-}\bar{n} & 0 & -\partial_z\bar{n} - \frac{\dot{n}}{2c}  
\end{array}\right).  	       
\end{eqnarray} 

In a source-free medium with time-independent $n$ and $\eta$ the time evolution equation for $\Psi$ is given by 
\begin{subequations} 
\label{SfreeTindep} 
\begin{equation}
\frac{\partial \Psi}{\partial t} = M\Psi, \quad  
M = \frac{c}{n}\,M_0 + M^\prime, 
\end{equation} 
\begin{equation}
M^\prime 
= \frac{c}{n}\left(\begin{array}{cc} 
({\mbox{\boldmath $\sigma$}}\cdot({\mbox{\boldmath $\nabla$}}\bar{n}))
\otimes{\mbox{\boldmath $1$}}_2 & -\left(({\mbox{\boldmath $\sigma$}}\cdot
({\mbox{\boldmath $\nabla$}}\bar{\eta}))\sigma_y\right)\otimes\sigma_y \\ 
-\left(({\mbox{\boldmath $\sigma$}}^*\cdot({\mbox{\boldmath $\nabla$}}\bar{\eta}))
\sigma_y\right)\otimes\sigma_y &  ({\mbox{\boldmath $\sigma$}}^*\cdot 
({\mbox{\boldmath $\nabla$}}\bar{n}))\otimes{\mbox{\boldmath $1$}}_2    
\end{array}\right), 	
\end{equation} 
\end{subequations} 
Explicitly writing, we have  
\begin{eqnarray}
M^\prime_{11} 
& = & \frac{c}{n}\left( 
\begin{array}{cccc} 
\partial_z\bar{n} & 0 & \partial_{-}\bar{n} & 0 \\ 
0 & \partial_z\bar{n} & 0 & \partial_{-}\bar{n} \\ 
\partial_+\bar{n} & 0 & -\partial_z\bar{n} & 0 \\ 
0 & \partial_+\bar{n} & 0 & -\partial_z\bar{n}  
\end{array}\right),  
\nonumber \\ 
M^\prime_{12} 
& = & \frac{c}{n}\left(  
\begin{array}{cccc} 
0 & -\partial_{-}\bar{\eta} & 0 & \partial_z\bar{\eta}  \\ 
\partial_{-}\bar{\eta} & 0 & -\partial_z\bar{\eta} & 0 \\ 
0 & \partial_z\bar{\eta} & 0 & \partial_+\bar{\eta} \\ 
-\partial_z\bar{\eta} & 0 & -\partial_+\bar{\eta} & 0  
\end{array}\right),  
\nonumber \\ 
M^\prime_{21} 
& = & \frac{c}{n}\left( 
\begin{array}{cccc} 
0 & -\partial_+\bar{\eta} & 0 & \partial_z\bar{\eta} \\ 
\partial_+\bar{\eta} & 0 & -\partial_z\bar{\eta} & 0 \\ 
0 & \partial_z\bar{\eta} & 0 & \partial_{-}\bar{\eta} \\ 
-\partial_z\bar{\eta} & 0 & -\partial_{-}\bar{\eta} & 0  
\end{array} \right),  
\nonumber \\  
M^\prime_{22} 		
& = & \frac{c}{n}\left( 
\begin{array}{cccc} 
\partial_z\bar{n} & 0 & \partial_{+}\bar{n} & 0 \\ 
0 & \partial_z\bar{n} & 0 & \partial_{+}\bar{n} \\ 
\partial_{-}\bar{n} & 0 & -\partial_z\bar{n} & 0 \\ 
0 & \partial_{-}\bar{n} & 0 & -\partial_z\bar{n} 
\end{array} \right).  	
\end{eqnarray} 
In a time-independent source-free inhomogeneous medium, $\mathcal{M}$ in 
Eq. (\ref{tiNCdFdt}), and $M$ in Eqs. (\ref{SfreeTindep}), are, in general, not antihermitian, as it is the case in a homogeneous medium.  Hence, the time evolution of $\Psi$, or $\mathcal{F}$, is, in general, not unitary for a time-independent source-free inhomogeneous medium (in this connection, see Koukoutsis et al., 2023; Vahala, et al., 2023).  
 
The eight dimensional matrix representation of the Maxwell equations for an inhomogeneous medium presented in (Khan, 2005) has been derived heuristically, by inspection of Eqs. (\ref{RSWMEq+}), (\ref{RSWMEq-}), and their generalizations in the case of an inhomogeneous medium.  The representation in (Khan, 2005) is equivalent to the new representation in Eq. (\ref{theEqn}), the fundamental time evolution equation for $\Psi$, through a unitary transformation.  To see this, 
let us write the matrix representation of the Maxwell equations for a linear inhomogeneous medium in (Khan, 2005) as 
\begin{subequations} 
\label{KhanRep}
\begin{eqnarray}
\frac{\partial\Psi_K}{\partial t} 
& = & M_K\Psi_K - W_K, \quad M_K = M_{0K} + M_K^\prime,  \nonumber \\    
\Psi_K 
& = & \left(\begin{array}{c} 
\psi^{+}_K \\ \psi^{-}_K 
\end{array}\right),  \quad 
W_K = \left(\begin{array}{l} 
{\mathfrak{j}}^{+}_K \\ {\mathfrak{j}}^{-}_K 
\end{array} \right),  
\end{eqnarray} 
\begin{equation} 
\psi^{+}_K 
= \left(\begin{array}{c} 
-F^{+}_x + iF^{+}_y \\ F^{+}_z \\ F^{+}_z \\ F^{+}_x + iF^{+}_y,    
\end{array} \right), \quad 
\psi^{-}_K 
= \left(\begin{array}{c} 
-F^{-}_x - iF^{-}_y \\ F^{-}_z \\ F^{-}_z \\ F^{-}_x - iF^{-}_y, 
\end{array} \right),  
\end{equation}  
\begin{equation} 
{\mathfrak{j}}^{+}_K 
= \frac{1}{\sqrt{2\epsilon}}  
\left(\begin{array}{c}
-J_x + iJ_y \\ J_z - v\rho \\ J_z + v\rho \\ J_x + iJ_y  
\end{array} \right),  \quad 
{\mathfrak{j}}^{-}_K 
=  \frac{1}{\sqrt{2\epsilon}}  
\left(\begin{array}{c}
-J_x - iJ_y \\ J_z - v\rho \\ J_z + v\rho \\ J_x - iJ_y  
\end{array} \right).
\end{equation} 
\begin{equation}
M_{0K} 
= -\frac{c}{n}
\left(\begin{array}{cc}
\left({\mbox{\boldmath $\sigma$}}\cdot{\mbox{\boldmath $\nabla$}}\right)
\otimes {\mbox{\boldmath $1$}}_2 & {\mbox{\boldmath $0$}}_4 \\ 
{\mbox{\boldmath $0$}}_4 & \left({\mbox{\boldmath $\sigma$}}^*\cdot
{\mbox{\boldmath $\nabla$}} \right)\otimes{\mbox{\boldmath $1$}}_2  
\end{array}\right), 	        
\end{equation} 	 
\begin{equation}
M_K^\prime 
= \frac{c}{n} 
\left(\begin{array}{cc} 
\left({\mbox{\boldmath $1$}}_2\otimes\left({\mbox{\boldmath $\sigma$}}\cdot
{\mbox{\boldmath $\nabla$}}\bar{n}\right)\right) 
- \frac{\dot{n}}{2c}{\mbox{\boldmath $1$}}_4 &  
\begin{array}{l} 
-\left(\sigma_y\otimes 
\left(\sigma_y\left({\mbox{\boldmath $\sigma$}}\cdot
{\mbox{\boldmath $\nabla$}}\bar{\eta}\right)\right)\right) \\ 
+ \frac{n\dot{\eta}}{2c\eta}(\sigma_y\otimes\sigma_y)
\end{array} \\ & \\  
\begin{array}{l} 
-\left(\sigma_y\otimes\left(\sigma_y
\left({\mbox{\boldmath $\sigma$}}^*\cdot{\mbox{\boldmath $\nabla$}} 
\bar{\eta}\right)\right)\right) \\ 
+ \frac{n\dot{\eta}}{2c\eta}(\sigma_y\otimes\sigma_y) 
\end{array} &   
\left({\mbox{\boldmath $1$}}_2\otimes\left({\mbox{\boldmath $\sigma$}}^*\cdot
{\mbox{\boldmath $\nabla$}}\bar{n}\right) \right) 
- \frac{\dot{n}}{2c}{\mbox{\boldmath $1$}}_4  
\end{array}\right). 
\end{equation}             
\end{subequations}  
Comparing $\Psi_K$, $W_K$, and $M_K$, with $\Psi$, $\mathfrak{J}$, and $M$, respectively, it is found that 
\begin{equation} 
\Psi = {\mathcal{S}}_K \Psi_K,  \quad  
{\mathfrak{J}} = {\mathcal{S}}_K W_K, \quad 
M = {\mathcal{S}}_K M_K {\mathcal{S}}_K^{\dagger},  
\end{equation} 
where 
\begin{equation} 
{\mathcal{S}}_K 
= \frac{1}{2}\left(\begin{array}{cccccccc}  
1 & 0 & 0 & 0 & 0 & 0 & 0 & 0 \\ 
0 & 0 & 1 & 0 & 0 & 0 & 0 & 0 \\ 
0 & 1 & 0 & 0 & 0 & 0 & 0 & 0 \\ 
0 & 0 & 0 & 1 & 0 & 0 & 0 & 0 \\ 
0 & 0 & 0 & 0 & 1 & 0 & 0 & 0 \\ 
0 & 0 & 0 & 0 & 0 & 0 & 1 & 0 \\ 
0 & 0 & 0 & 0 & 0 & 1 & 0 & 0 \\ 
0 & 0 & 0 & 0 & 0 & 0 & 0 & 1  
\end{array}\right).        
\end{equation}  
Note that $S_K$ is a permutation matrix and is unitary.  If we substitute ${\mathcal{S}}_K^{\dagger}\Psi$, ${\mathcal{S}}_K^{\dagger}{\mathfrak{J}}$, and ${\mathcal{S}}_K^{\dagger} M {\mathcal{S}}_K$, respectively, for $\Psi_K$, $W_K$, and $M_K$ in Eq. (\ref{KhanRep}) and multiply throughout from left by ${\mathcal{S}}_K$ then we get Eq. (\ref{theEqn}).  Note that $\Psi$ and $\Psi_K$ are identical except for a constant factor of $1/2$.  The matrix elements of $U$ in 
Eq. (\ref{Uchoice}) and the extra $-i$ multiplying $U^*$ in the lower diagonal block of $T$ in Eq. (\ref{Tdefn}) have been chosen in the particular way only to make $\Psi$ identical to $\Psi_K$ except for an overall constant factor.  Other consistent choices will lead to different expressions for $\Psi$.  However, the final results for the physical fields will be the same in any problem when the fields are retrieved from $\Psi$ by the corresponding inverse transformations like in Eq. (\ref{InvTransf}).  
 
\section{Applications} 

\subsection{Electromagnetic wave propagation in a homogeneous medium}
 
For a time-independent homogeneous source-free medium characterized by constant $n$ and $\eta$ 
Eqs. (\ref{theEqn}) become 
\begin{equation}
\frac{\partial\Psi}{\partial t} 
= -v\left(\begin{array}{cc} 
{\mbox{\boldmath $\Sigma$}}\cdot{\mbox{\boldmath $\nabla$}} & 
{\mbox{\boldmath $0$}} \\ 
{\mbox{\boldmath $0$}} & {\mbox{\boldmath $\Sigma$}}^*\cdot
{\mbox{\boldmath $\nabla$}}  
\end{array}\right)\Psi,    
\label{Sfreepsiteq}
\end{equation}
where $v = c/n$.  This equation can be written as a pair of equations,   
\begin{subequations} 
\label{SfreeEqns4vcum} 	
\begin{equation} 
\frac{\partial \psi^{+}}{\partial t} 
= - v({\mbox{\boldmath $\Sigma$}}\cdot{\mbox{\boldmath $\nabla$}})\psi^{+}, 
\label{SfreeEqns4vcum-a}
\end{equation} 
\begin{equation}  
\frac{\partial\psi^{-}}{\partial t} 
= -v({\mbox{\boldmath $\Sigma$}}^*\cdot{\mbox{\boldmath $\nabla$}})\psi^{-}.     
\label{SfreeEqns4vcum-b}
\end{equation} 
\end{subequations}   
Here, $\psi^{+}$ and $\psi^{-}$ are the four upper components and the four lower components, respectively, of the  $8$-component $\Psi$ defined in 
Eq. (\ref{psi+-j+-defn}).  With $v$ being constant and 
${\mbox{\boldmath $\Sigma$}}\cdot{\mbox{\boldmath $\nabla$}}$ and 
${\mbox {\boldmath $\Sigma$}}^*\cdot{\mbox{\boldmath $\nabla$}}$ being time-independent, from Eqs. (\ref{SfreeEqns4vcum}) we get 
\begin{equation}
{\nabla}^2\psi^{\pm} 
- \frac{1}{v^2}\frac{\partial^2\psi^{\pm}}{\partial t^2} = 0.       
\end{equation} 
This implies,  
\begin{equation}
{\nabla}^2{\mbox{\boldmath $F$}}^{\pm} - \frac{1}{v^2}
\frac{\partial^2{\mbox{\boldmath $F$}}^{\pm}}{\partial t^2} = 0, 
\label{Helmholtz4F}
\end{equation} 
since each component of $\psi^{\pm}$ is a linear combination of the components of 
${\mbox{\boldmath $F$}}^{\pm}$.  This is same as the Helmholtz equations in 
Eq. (\ref{Helmholtz}) since ${\mbox{\boldmath $F$}}^{+}$ and 
${\mbox{\boldmath $F$}}^{-}$ are linear combinations of ${\mbox{\boldmath $E$}}$ and ${\mbox{\boldmath $B$}}$.  

As is usual in optics, we shall specify any electromagnetic wave by the complex notation with the real part representing the physical field.  Let  $\left\{\widehat{\mbox{\boldmath $x$}},\widehat{\mbox{\boldmath $y$}},
\widehat{\mbox{\boldmath $z$}}\right\}$ denote the unit vectors in an $(x,y,z)$-Cartesian coordinate system.  Let us consider a plane electromagnetic wave of circular frequency $\omega$ and wave vector $\mbox{\boldmath{$k$}} 
= k_x\widehat{\mbox{\boldmath $x$}} + k_y\widehat{\mbox{\boldmath $y$}} + k_z\widehat{\mbox{\boldmath $z$}}$ propagating in a medium without any charges or currents.  The electromagnetic field of the wave may, in general, be taken to be 
\begin{equation} 
{\mbox{\boldmath $E$}} 
= {\mbox{\boldmath $E$}}_0 \exp\{i\left({\mbox{\boldmath $k$}}\cdot
{\mbox{\boldmath $r$}}-{\omega}t\right)\}, \quad 
{\mbox{\boldmath $B$}} 
= {\mbox{\boldmath $B$}}_0 \exp\{i\left({\mbox{\boldmath $k$}}\cdot
{\mbox{\boldmath $r$}}-{\omega}t\right)\},   
\end{equation} 
where ${\mbox{\boldmath $E$}}_0$ and ${\mbox{\boldmath $B$}}_0$ are the constant complex amplitude vectors including the phase. Then, the plane wave is associated with 
\begin{eqnarray}
{\mbox{\boldmath $F$}}^{\pm} 
& = & \frac{1}{\sqrt{2}} \left(\sqrt{\epsilon}{\mbox{\boldmath $E$}}_0 
\pm \frac{i}{\sqrt{\mu}}{\mbox{\boldmath $B$}}_0\right) 
\exp\{i\left({\mbox{\boldmath $k$}}\cdot
{\mbox{\boldmath $r$}}-{\omega}t\right)\}  
\nonumber \\        
& = & {\mbox{\boldmath $F$}}^{\pm}_0 
\exp\{i\left({\mbox{\boldmath $k$}}\cdot
{\mbox{\boldmath $r$}}-{\omega}t\right)\}, 
\label{planewaveF} 
\end{eqnarray} 
which have to satisfy Eq. (\ref{SfreeRSWMEq}) and Eq. (\ref{Helmholtz4F}).  Substituting this expression for $\mbox{\boldmath $F$}^{\pm}$ in 
Eq. (\ref{Helmholtz4F}) we get the well known dispersion relation for electromagnetic waves in a source-free medium of constant $n$,  
\begin{equation} 
\omega = v\sqrt{k_x^2 + k_y^2 + k_z^2} = vk,    
\label{dispersion}
\end{equation}
with $k$ as the magnitude of the wave vector equal to $2\pi n/\lambda$ where $\lambda$ is the wavelength in vacuum.  Substitution of the expression in 
Eq. (\ref{planewaveF}) for ${\mbox{\boldmath $F$}}^{\pm}$ in Eq. (\ref{SfreeRSWMEq}) leads to the conditions 
\begin{equation}
{\mbox{\boldmath $k$}}\cdot{\mbox{\boldmath $F$}}^{\pm}_0 = 0, 	\quad
k{\mbox{\boldmath $F$}}^{\pm}_0 
= \pm i\left({\mbox{\boldmath $k$}}\times{\mbox{\boldmath $F$}}^{\pm}_0\right),	
\label{transversality} 
\end{equation} 
which imply the well known relations between $\mbox{\boldmath $E$}$,
$\mbox{\boldmath $B$}$, and $\mbox{\boldmath $k$}$ of the plane wave 
\begin{eqnarray}
{\mbox{\boldmath $k$}}\cdot{\mbox{\boldmath $E$}}_0 
& = & 0,  \quad 
{\mbox{\boldmath $k$}}\cdot{\mbox{\boldmath $B$}}_0 = 0,  
\nonumber \\ 
\widehat{\mbox{\boldmath $k$}}\times{\mbox{\boldmath $E$}}_0  
& = & v{\mbox{\boldmath $B$}}_0, \quad 
\widehat{\mbox{\boldmath $k$}}\times{v\mbox{\boldmath $B$}}_0  
= -{\mbox{\boldmath $E$}}_0,   
\end{eqnarray}  
where $\widehat{\mbox{\boldmath $k$}} = {\mbox{\boldmath $k$}}/k$ is the unit vector in the direction of propagation of the plane wave.  

Equation (\ref{SfreeEqns4vcum-a}) for $\psi^{+}$ can be readily integrated with respect to time to give 
\begin{eqnarray} 
\psi^{+}(t) 
& = & \exp\{-v\left(t-t_0\right)\,\left({\mbox{\boldmath $\Sigma$}}\cdot 
{\mbox{\boldmath $\nabla$}}\right)\}\psi^{+}\left(t_0\right)  
\nonumber \\ 
& = & \left(1- v\left(t-t_0\right)
\left({\mbox{\boldmath $\Sigma$}}\cdot{\mbox{\boldmath $\nabla$}}\right) 
+ \frac{\left(v\left(t-t_0\right)\right)^2}{2!}\nabla^2 \right. 
\nonumber \\ 
&   & \qquad - \frac{\left(v\left(t-t_0\right)\right)^3}{3!}\nabla^2
\left({\mbox{\boldmath $\Sigma$}}\cdot{\mbox{\boldmath $\nabla$}}\right) 
+ \frac{\left(v\left(t-t_0\right)\right)^4}{4!}\nabla^4 
\nonumber \\ 
&   & \qquad \left. -  \frac{\left(v\left(t-t_0\right)\right)^5}{5!}\nabla^4 
\left({\mbox{\boldmath $\Sigma$}}\cdot{\mbox{\boldmath $\nabla$}}\right)  
+ \cdots \right)\psi^{+}\left(t_0\right)  
\nonumber \\ 
& = & \left(\sum_{j=0}^{\infty}
\frac{\left(v\left(t-t_0\right)\right)^{2j}}{(2j)!} \nabla^{2j} \right. 
\nonumber \\ 
&   & \qquad \left. -\sum_{j=0}^{\infty} \frac{\left(v\left(t-t_0\right)\right)^{2j+1}}{(2j+1)!}
\nabla^{2j} \left({\mbox{\boldmath $\Sigma$}}\cdot
{\mbox{\boldmath $\nabla$}}\right)\right)\psi^{+}\left(t_0\right).      
\label{psi+t0+t}
\end{eqnarray} 
For a plane wave we have 
\begin{eqnarray}
\psi^{+}\left(t_0\right) 
& = & \frac{1}{2}\left(\begin{array}{c}
-F^{+}_{0x}+iF^{+}_{0y} \\ F^{+}_{0z} \\ F^{+}_{0z} \\ F^{+}_{0x}+iF^{+}_{0y} 
\end{array}\right) 
\exp\{i\left({\mbox{\boldmath $k$}}\cdot{\mbox{\boldmath $r$}}
-{\omega}t_0\right)\}  
\nonumber \\     
& = & \psi^{+}_0 \exp\{i\left({\mbox{\boldmath $k$}}\cdot 
{\mbox{\boldmath $r$}}-{\omega}t_0\right)\}.       
\end{eqnarray}  
Now, we observe that 
\begin{equation} 
\nabla^{2j}\exp\{i\left({\mbox{\boldmath $k$}}\cdot{\mbox{\boldmath $r$}}
-{\omega}t_0\right)\} 
= (-1)^jk^{2j}\exp\{i\left({\mbox{\boldmath $k$}}\cdot{\mbox{\boldmath $r$}}
-{\omega}t_0\right)\},  
\label{nabla2} 
\end{equation} 
and 
\begin{eqnarray}
\left({\mbox{\boldmath $\Sigma$}}\cdot{\mbox{\boldmath $\nabla$}}\right)
\psi^{+}\left(t_0\right) 
& = & \left(\begin{array}{cc} 
{\mbox{\boldmath $\sigma$}}\cdot{\mbox{\boldmath $\nabla$}} & 
{\mbox{\boldmath $0$}} \\ 
{\mbox{\boldmath $0$}} & 
{\mbox{\boldmath $\sigma$}}\cdot{\mbox{\boldmath $\nabla$}} 
\end{array}\right) 
\psi^{+}_0 \exp\{i\left({\mbox{\boldmath $k$}}\cdot
{\mbox{\boldmath $r$}}-{\omega}t_0\right)\}  
\nonumber \\  
& = & ik\psi^{+}_0 \exp\{i\left({\mbox{\boldmath $k$}}\cdot{\mbox{\boldmath $r$}}
-{\omega}t_0\right)\} 
= ik\psi^{+}\left(t_0\right),  
\label{nabla1} 
\end{eqnarray}
where we have used the transversality conditions on ${\mbox{\boldmath $F$}}^{+}_0$ in Eq. (\ref{transversality}).  In view of Eqs. (\ref{dispersion}), (\ref{nabla2}), and (\ref{nabla1}), Eq. (\ref{psi+t0+t}) becomes 
\begin{eqnarray} 
\psi^{+}(t) 
& = & \left(\sum_{j=0}^{\infty}\frac{(-1)^j
\left({\omega}\left(t-t_0\right)\right)^{2j}}{(2j)!} 
-i\sum_{j=0}^{\infty} \frac{(-1)^j
\left({\omega}\left(t-t_0\right)\right)^{2j+1}}{(2j+1)!}\right) 
\nonumber \\ 
&   & \qquad \times \psi^{+}_0\exp\{i\left({\mbox{\boldmath $k$}}\cdot
{\mbox{\boldmath $r$}}-{\omega}t_0\right)\}  
\nonumber \\   
& = & \exp\{-i{\omega}\left(t-t_0\right)\}\psi^{+}_0 
\exp\{i\left({\mbox{\boldmath $k$}}\cdot
{\mbox{\boldmath $r$}}-{\omega}t_0\right)\}  
\nonumber \\ 
& = & \psi^{+}_0\exp\{i\left({\mbox{\boldmath $k$}}
\cdot{\mbox{\boldmath $r$}}-{\omega}t\right)\}.      
\label{psi+t} 
\end{eqnarray}  
Similarly, for 
\begin{eqnarray} 
\psi^{-}(t) 
& = & \frac{1}{2}
\left(\begin{array}{c}
-F^{-}_{0x}-iF^{-}_{0y} \\ F^{-}_{0z} \\ F^{-}_{0z} \\  F^{-}_{0x}-iF^{-}_{0y} \end{array}\right)
\exp\{i\left({\mbox{\boldmath $k$}}\cdot
{\mbox{\boldmath $r$}}-{\omega}t\right)\}  
\nonumber \\     
& = & \psi^{-}_0 \exp\{i\left({\mbox{\boldmath $k$}}\cdot
{\mbox{\boldmath $r$}}-{\omega}t\right)\},  
\end{eqnarray} 
we get 
\begin{equation}
\exp\{-v\left(t-t_0\right)\,\left({\mbox{\boldmath $\Sigma$}}^* 
\cdot {\mbox{\boldmath $\nabla$}}\right)\} \psi^{-}\left(t_0\right) 
= \psi^{-}(t), 
\label{psi-t0+t}
\end{equation} 
starting with Eq. (\ref{SfreeEqns4vcum}).  Since any electromagnetic wave packet is a linear combination of plane waves, it follows from Eqs. (\ref{psi+t0+t}),  (\ref{psi+t}), and (\ref{psi-t0+t}), that Eq. (\ref{Sfreepsiteq}) can be used to study electromagnetic wave propagation problems for a source-free medium of constant $\epsilon$ and $\mu$. 

\subsection{Analogy with quantum mechanics}

It should be noted that for a source-free inhomogeneous medium with time-independent $n$ and $\eta$ the fundamental time evolution equation in Eq. (\ref{SfreeTindep}) has the structure of the Schr\"{o}dinger equation for a quantum mechanical system, namely,  
\begin{subequations}  
\label{Hamiltonian} 
\begin{equation}
\frac{\partial\Psi}{\partial t} = \widehat{H}\Psi,  \quad   
\widehat{H} = \widehat{H}_0 + \widehat{H}^\prime,  	
\end{equation}	
\begin{equation} 
\widehat{H}_0 
= -\frac{c}{n}\left(\begin{array}{cc} 
{\mbox{\boldmath $\Sigma$}}\cdot{\mbox{\boldmath $\nabla$}} & 
{\mbox{\boldmath $0$}}_4 \\ 
{\mbox{\boldmath $0$}}_4 & 
{\mbox{\boldmath $\Sigma$}}^*\cdot{\mbox{\boldmath $\nabla$}}  
\end{array}\right), 	
\end{equation}
\begin{equation}
\widehat{H}^\prime 
= \frac{c}{n}\left(\begin{array}{cc} 
({\mbox{\boldmath $\sigma$}}\cdot({\mbox{\boldmath $\nabla$}}\bar{n})) 
\otimes{\mbox{\boldmath $1$}}_2 & 
-\left(({\mbox{\boldmath $\sigma$}}\cdot
({\mbox{\boldmath $\nabla$}}\bar{\eta}))\sigma_y\right)\otimes\sigma_y 
\\ & \\ 
-\left(({\mbox{\boldmath $\sigma$}}^*\cdot
({\mbox{\boldmath $\nabla$}}\bar{\eta}))\sigma_y\right)\otimes\sigma_y  
& ({\mbox{\boldmath $\sigma$}}^*\cdot({\mbox{\boldmath $\nabla$}}\bar{n}))
\otimes{\mbox{\boldmath $1$}}_2     
\end{array}\right),    	
\end{equation} 
\end{subequations} 
where the ``Hamiltonian'' $\widehat{H}$ is split into a core, or unperturbed, Hamiltonian $\widehat{H}_0$ and a perturbation Hamiltonian $\widehat{H}^\prime$.  
If we drop the perturbation part, in the zeroth order approximation, then we have 
\begin{equation} 
\frac{\partial\Psi}{\partial t} 
= \widehat{H}_0\Psi 
= -\frac{c}{n}
\left(\begin{array}{cc} 
{\mbox{\boldmath $\Sigma$}}\cdot{\mbox{\boldmath $\nabla$}} & 
{\mbox{\boldmath $0$}}_4 \\ 
{\mbox{\boldmath $0$}}_4 & 
{\mbox{\boldmath $\Sigma$}}^*\cdot{\mbox{\boldmath $\nabla$}}  
\end{array}\right)\Psi.   
\end{equation} 
Since $\widehat{H}_0$ is time-independent, we have 
\begin{eqnarray} 
\frac{\partial^2\Psi}{\partial t^2} 
& = & \widehat{H}_0^2\Psi  
\nonumber \\ 
& = & \frac{c}{n}
\left(\begin{array}{cc} 
{\mbox{\boldmath $\Sigma$}}\cdot{\mbox{\boldmath $\nabla$}} & 
{\mbox{\boldmath $0$}}_4 \\ 
{\mbox{\boldmath $0$}}_4 & 
{\mbox{\boldmath $\Sigma$}}^*\cdot{\mbox{\boldmath $\nabla$}}  
\end{array}\right) 
\frac{c}{n}
\left(\begin{array}{cc} 
{\mbox{\boldmath $\Sigma$}}\cdot{\mbox{\boldmath $\nabla$}} & 
{\mbox{\boldmath $0$}}_4 \\ 
{\mbox{\boldmath $0$}}_4 &  
{\mbox{\boldmath $\Sigma$}}^*\cdot{\mbox{\boldmath $\nabla$}}  
\end{array}\right)\Psi.                                
\end{eqnarray}  
Note that 
\begin{eqnarray} 
\frac{1}{n}({\mbox{\boldmath $\sigma$}}\cdot{\mbox{\boldmath $\nabla$}})
\frac{1}{n}({\mbox{\boldmath $\sigma$}}\cdot{\mbox{\boldmath $\nabla$}})   
& = & \frac{1}{n}
\left(\begin{array}{cc} 
\partial_z & \partial_{-} \\ \partial_{+} & -\partial_z 
\end{array}\right)
\frac{1}{n}
\left(\begin{array}{cc}	                 
\partial_z & \partial_{-} \\ \partial_{+} & -\partial_z 
\end{array}\right)  
\nonumber \\ 
& = & \left(\begin{array}{cc}
m_{11} & m_{12} \\ m_{21} & m_{22} 
\end{array}\right),       
\end{eqnarray} 
with  
\begin{eqnarray}
m_{11} 
& = & \frac{1}{n^2}\nabla^2-\frac{1}{n^3}(({\mbox{\boldmath $\nabla$}}n) 
\cdot{\mbox{\boldmath $\nabla$}})
-\frac{i}{n^3}\left(({\mbox{\boldmath $\nabla$}}n) 
\times {\mbox{\boldmath $\nabla$}}\right)_z,  
\nonumber \\ 
m_{12} & = & \frac{1}{n^3}\left(\left(\partial_z n\right)\partial_{-} 
+ \left(\partial_{+}n\right)\partial_z\right),  \quad 
m_{21} = \frac{1}{n^3}\left(\left(\partial_z n\right)\partial_{+} 
+ \left(\partial_{-}n\right)\partial_z\right),  
\nonumber \\  
m_{22} & = & \frac{1}{n^2}\nabla^2-\frac{1}{n^3}(({\mbox{\boldmath $\nabla$}}n) 
\cdot {\mbox{\boldmath $\nabla$}}) 
+ \frac{i}{n^3}\left(({\mbox{\boldmath $\nabla$}}n) 
\times {\mbox{\boldmath $\nabla$}}\right)_z.  
\end{eqnarray} 
Discarding the terms $\sim 1/n^3$, we can take 
\begin{equation}
\frac{\partial^2\Psi}{\partial t^2} \approx \frac{c^2}{n^2}\nabla^2\Psi.    
\end{equation} 
This leads to the scalar wave equation 
\begin{equation} 
\left(\nabla^2-\frac{n(\mbox{\boldmath $r$})^2}{c^2}
\frac{\partial^2}{\partial t^2}\right)\Psi(\mbox{\boldmath $r$},t) = 0,    
\end{equation} 
the basis of the Helmholtz scalar wave optics.  To go beyond the scalar wave optics and understand the propagation of electromagnetic vector waves in an inhomogeneous medium, including polarization, one has to study the time evolution in 
Eq. (\ref{Hamiltonian}) taking into account the spatial variations in $n$ and $\eta$.  The analogy with quantum mechanics should help study this equation using suitable techniques.    

\subsection{Maxwell vector wave optics}

Let us now consider the propagation of a monochromatic electromagnetic beam of circular frequency $\omega$ along the $z$-axis in an inhomogeneous source-free time-independent medium.  This system is best studied in a representation provided by the following unitary transformation:  
\begin{equation}
\Phi(\mbox{\boldmath $r$},t) 
= {\mathcal{S}}_{\phi}\Psi(\mbox{\boldmath $r$},t),  \quad 
{\mathcal{S}}_{\phi}  
= \left(\begin{array}{cccccccc} 
1 & 0 & 0 & 0 & 0 & 0 & 0 & 0 \\ 
0 & 0 & 1 & 0 & 0 & 0 & 0 & 0 \\ 
0 & 0 & 0 & 0 & 1 & 0 & 0 & 0 \\ 
0 & 0 & 0 & 0 & 0 & 0 & 1 & 0 \\
0 & 1 & 0 & 0 & 0 & 0 & 0 & 0 \\ 
0 & 0 & 0 & 1 & 0 & 0 & 0 & 0 \\ 
0 & 0 & 0 & 0 & 0 & 1 & 0 & 0 \\
0 & 0 & 0 & 0 & 0 & 0 & 0 & 1       
\end{array}\right).  
\label{phitransf}
\end{equation} 
This makes 
\begin{equation} 
\Phi 
= \left(\begin{array}{c}
\phi^{+} \\ \phi^{-}
\end{array}\right),  \quad 
\phi^{+} 
= \frac{1}{2} 
\left(\begin{array}{c} 
-F^{+}_x + iF^{+}_y \\ F^{+}_z \\ -F_x^{-} - iF_y^{-} \\ F_z^{-} 
\end{array}\right),  \quad 
\phi^{-} 
= \frac{1}{2} 
\left(\begin{array}{c} 
F^{+}_z \\ F^{+}_x + iF^{+}_y \\ F_z^{-} \\ F_x^{-} - iF_y^{-}  
\end{array}\right).  
\end{equation} 
The fields $\mbox{\boldmath $E$}$ and $\mbox{\boldmath $B$}$ can be recovered from $\Phi$ using the inverse transformation 
\begin{equation}
\mathcal{F} = \mathbb{T}^\dagger\mathcal{S}_{\phi}^\dagger\Phi.  	
\end{equation}  
The purpose of the transformation in Eq. (\ref{phitransf}) is to change the coefficient matrix of $\partial_z$ in $M$ in Eq. (\ref{SfreeTindep}), namely 
$\mbox{\boldmath $1$}_4 \otimes \sigma_z$, to 
$\sigma_z \otimes \mbox{\boldmath $1$}_4$.  It can be easily verified that 
\begin{equation}
{\mathcal{S}}_{\phi}\left(\mbox{\boldmath $1$}_4 \otimes \sigma_z\right)
{\mathcal{S}}_{\phi}^{\dagger} 
= \sigma_z \otimes \mbox{\boldmath $1$}_4.  
\end{equation} 
As a result, the time evolution equation for $\Phi$, as obtained from 
Eq. (\ref{SfreeTindep}), is 
\begin{subequations} 
\label{phirepeqn} 
\begin{equation}
\frac{\partial\Phi}{\partial t} 
= \left(\mathcal{S}_{\phi} M \mathcal{S}_{\phi}^\dagger\right)\Phi 
= \mathbb{M}\Phi,  \quad  
\mathbb{M} = \mathbb{M}_0 + \mathbb{M}^\prime,  
\end{equation}
\begin{equation} 
\mathbb{M}_0 
= -\frac{c}{n}
\left(\begin{array}{cccc}
{\mbox{\boldmath $1$}}_2\partial_z & {\mbox{\boldmath $0$}}_2 & 
{\mbox{\boldmath $1$}}_2\partial_{-} & {\mbox{\boldmath $0$}}_2 \\
{\mbox{\boldmath $0$}}_2 & {\mbox{\boldmath $1$}}_2\partial_z & 
{\mbox{\boldmath $0$}}_2 & {\mbox{\boldmath $1$}}_2\partial_{+} \\ 
{\mbox{\boldmath $1$}}_2\partial_{+} & {\mbox{\boldmath $0$}}_2 & 
-{\mbox{\boldmath $1$}}_2\partial_{z} & {\mbox{\boldmath $0$}}_2 \\
{\mbox{\boldmath $0$}}_2 & {\mbox{\boldmath $1$}}_2\partial_{-} & 
{\mbox{\boldmath $0$}}_2 & -{\mbox{\boldmath $1$}}_2\partial_z 
\end{array}\right),   	
\end{equation}
\begin{equation}
\mathbb{M}^\prime 
= \frac{c}{n}\left(\begin{array}{cccc}  
{\mbox{\boldmath $\sigma$}}\cdot{\mbox{\boldmath $\nabla$}}\bar{n} & 
{\mbox{\boldmath $0$}}_2 & {\mbox{\boldmath $0$}}_2 & 
-i\sigma_y({\mbox{\boldmath $\sigma$}}^*\cdot
{\mbox{\boldmath $\nabla$}}\bar{\eta}) \\ 
{\mbox{\boldmath $0$}}_2 & {\mbox{\boldmath $\sigma$}}^*\cdot
{\mbox{\boldmath $\nabla$}}\bar{n} & -i\sigma_y({\mbox{\boldmath $\sigma$}}\cdot
{\mbox{\boldmath $\nabla$}}\bar{\eta}) & {\mbox{\boldmath $0$}}_2 \\ 
{\mbox{\boldmath $0$}}_2 & i\sigma_y({\mbox{\boldmath $\sigma$}}^*\cdot
{\mbox{\boldmath $\nabla$}}\bar{\eta}) & {\mbox{\boldmath $\sigma$}}\cdot
{\mbox{\boldmath $\nabla$}}\bar{n} & {\mbox{\boldmath $0$}}_2 \\ 
i\sigma_y({\mbox{\boldmath $\sigma$}}\cdot
{\mbox{\boldmath $\nabla$}}\bar{\eta}) & {\mbox{\boldmath $0$}}_2 & 
{\mbox{\boldmath $0$}}_2 & {\mbox{\boldmath $\sigma$}}^*\cdot
{\mbox{\boldmath $\nabla$}}\bar{n}  
\end{array}\right).  
\end{equation}
\end{subequations} 
Note that the coefficient matrix of $\partial_z$ in Eq. (\ref{phirepeqn}) is 
$\sigma_z \otimes \mbox{\boldmath $1$}_4$ as required.  Of course, the other matrix elements have also changed in $\mathbb{M}$, but we are not concerned about them. 

We can take the field associated with the monochromatic beam to be 
\begin{equation} 
\mbox{\boldmath $E$}(\mbox{\boldmath $r$},t) 
= \bar{\mbox{\boldmath $E$}}({\mbox{\boldmath $r$}}_\perp,z)\exp\{-i\omega t\},  
\qquad  
\mbox{\boldmath $B$}(\mbox{\boldmath $r$},t) 
= \bar{\mbox{\boldmath $B$}}({\mbox{\boldmath $r$}}_\perp,z)\exp\{-i\omega t\},  
\end{equation} 
where $\left(\bar{\mbox{\boldmath $E$}}({\mbox{\boldmath $r$}}_\perp,z),
\bar{\mbox{\boldmath $B$}}({\mbox{\boldmath $r$}}_\perp,z)\right)$ is the field in the $xy$-plane at the point $z$ on the axis of the beam at $t = 0$.  Correspondingly we have 
\begin{equation}
\Phi(\mbox{\boldmath $r$},t) 
= \bar{\Phi}({\mbox{\boldmath $r$}}_\perp,z) \exp\{-i\omega t\}   
= \left(\begin{array}{c} 
\bar{\phi}^{+}({\mbox{\boldmath $r$}}_\perp,z) \\
\bar{\phi}^{-}({\mbox{\boldmath $r$}}_\perp,z)   
\end{array}\right) \exp\{-i\omega t\}.     
\end{equation} 
Substituting ${\Phi}(\mbox{\boldmath $r$},t)$ in Eq. (\ref{phirepeqn}), canceling $\exp\{-i\omega t\}$ on both sides of the resulting equation, and rearranging the terms, we get 
\begin{eqnarray}
&   & \frac{c}{n}
\left(\begin{array}{cr}
{\mbox{\boldmath $1$}}_4\partial_z & {\mbox{\boldmath $0$}}_4 \\ 
{\mbox{\boldmath $0$}}_4 & -{\mbox{\boldmath $1$}}_4\partial_z  
\end{array}\right)
\left(\begin{array}{c} 
\bar{\phi}^{+}({\mbox{\boldmath $r$}}_\perp,z) \\ 
\bar{\phi}^{-}({\mbox{\boldmath $r$}}_\perp,z)   
\end{array}\right)  \nonumber \\ 
&   & \qquad 
= \left[i\omega{\mbox{\boldmath $1$}}_8
-\frac{c}{n}\left(\begin{array}{cccc}
{\mbox{\boldmath $0$}}_2 & {\mbox{\boldmath $0$}}_2 & 
{\mbox{\boldmath $1$}}_2\partial_{-} & {\mbox{\boldmath $0$}}_2 \\
{\mbox{\boldmath $0$}}_2 & {\mbox{\boldmath $0$}}_2 & 
{\mbox{\boldmath $0$}}_2 & {\mbox{\boldmath $1$}}_2\partial_{+} \\ 
{\mbox{\boldmath $1$}}_2\partial_{+} & {\mbox{\boldmath $0$}}_2 & 
{\mbox{\boldmath $0$}}_2 & {\mbox{\boldmath $0$}}_2 \\
{\mbox{\boldmath $0$}}_2 & {\mbox{\boldmath $1$}}_2\partial_{-} & 
{\mbox{\boldmath $0$}}_2 & {\mbox{\boldmath $0$}}_2   
\end{array}\right) + \mathbb{M}^\prime\right] 
\left(\begin{array}{c} 
\bar{\phi}^{+}({\mbox{\boldmath $r$}}_\perp,z) \\
\bar{\phi}^{-}({\mbox{\boldmath $r$}}_\perp,z)   
\end{array}\right).  
\nonumber \\ 
   &   &   
\label{barphieq}
\end{eqnarray}
Let us define the eight dimensional analog of Dirac's $\beta$ matrix as 
\begin{equation}
\mathcal{B} 
= \sigma_z\otimes{\mbox{\boldmath $1$}}_4  
= \left(\begin{array}{cr} 
{\mbox{\boldmath $1$}}_4 &  {\mbox{\boldmath $0$}}_4 \\ 
{\mbox{\boldmath $0$}}_4 & -{\mbox{\boldmath $1$}}_4 
\end{array}\right).     
\end{equation}
For a monochromatic beam $\lambda = 2\pi c/\omega$ is the wavelength while traveling in vacuum and $\kappa = 2\pi/\lambda = \omega/c$ is the magnitude of the corresponding wave vector.  If we  write $\lambda/2\pi = \bar{\lambda}$, then $c/\omega = \bar{\lambda} = 1/\kappa$.  Multiplying both sides of 
Eq. (\ref{barphieq}) from left by $i(n/\omega)\mathcal{B}$ and rearranging the terms we get  
\begin{subequations} 
\label{beamoptH} 
\begin{eqnarray} 
i\bar{\lambda} \frac{\partial}{\partial z} 
\left(\begin{array}{c} 
\bar{\phi}^{+}({\mbox{\boldmath $r$}}_\perp,z) \\
\bar{\phi}^{-}({\mbox{\boldmath $r$}}_\perp,z)   
\end{array}\right) 
& = & \widehat{\mathcal{H}}
\left(\begin{array}{c} 
\bar{\phi}^{+}({\mbox{\boldmath $r$}}_\perp,z) \\
\bar{\phi}^{-}({\mbox{\boldmath $r$}}_\perp,z)   
\end{array}\right),  \nonumber \\  
\widehat{\mathcal{H}} 
& = & -n_0\mathcal{B} + \widehat{\mathcal{E}} + \widehat{\mathcal{O}},  	
\end{eqnarray} 
\begin{equation}
\widehat{\mathcal{E}} 
= -\left( n - n_0 \right)\mathcal{B} 
+ i\bar{\lambda} 
\left(\begin{array}{cccc}  
{\mbox{\boldmath $\sigma$}}\cdot{\mbox{\boldmath $\nabla$}}\bar{n} & 
{\mbox{\boldmath $0$}}_2 & {\mbox{\boldmath $0$}}_2 & 
{\mbox{\boldmath $0$}}_2 \\ {\mbox{\boldmath $0$}}_2 & 
{\mbox{\boldmath $\sigma$}}^*\cdot{\mbox{\boldmath $\nabla$}}\bar{n} & 
{\mbox{\boldmath $0$}}_2 & {\mbox{\boldmath $0$}}_2 \\ 
{\mbox{\boldmath $0$}}_2 & {\mbox{\boldmath $0$}}_2 & 
-{\mbox{\boldmath $\sigma$}}\cdot{\mbox{\boldmath $\nabla$}}\bar{n} & 
{\mbox{\boldmath $0$}}_2 \\ 
{\mbox{\boldmath $0$}}_2 & {\mbox{\boldmath $0$}}_2 & 
{\mbox{\boldmath $0$}}_2 & 
-{\mbox{\boldmath $\sigma$}}^*\cdot{\mbox{\boldmath $\nabla$}}\bar{n}  
\end{array}\right), 
\end{equation}
\begin{equation}
\widehat{\mathcal{O}}  
= \left(\begin{array}{cccc}  
{\mbox{\boldmath $0$}}_2 & {\mbox{\boldmath $0$}}_2 & 
{\mbox{\boldmath $1$}}_2\widehat{\wp}_{-} & 
\begin{array}{l}
\bar{\lambda}\sigma_y\times \\ ({\mbox{\boldmath $\sigma$}}^*\cdot
{\mbox{\boldmath $\nabla$}}\bar{\eta}) 
\end{array} \\ 
{\mbox{\boldmath $0$}}_2 & {\mbox{\boldmath $0$}}_2 &	
\begin{array}{l}
\bar{\lambda}\sigma_y\times \\ ({\mbox{\boldmath $\sigma$}}\cdot
{\mbox{\boldmath $\nabla$}}\bar{\eta}) 
\end{array} & 
{\mbox{\boldmath $1$}}_2\widehat{\wp}_{+} \\ 
-{\mbox{\boldmath $1$}}_2\widehat{\wp}_{+} & 
\begin{array}{l} 
\bar{\lambda}\sigma_y\times \\ ({\mbox{\boldmath $\sigma$}}^*\cdot
{\mbox{\boldmath $\nabla$}}\bar{\eta}) 
\end{array} & 
{\mbox{\boldmath $0$}}_2 & {\mbox{\boldmath $0$}}_2 \\ 
\begin{array}{l} 
\bar{\lambda}\sigma_y\times \\ ({\mbox{\boldmath $\sigma$}}\cdot
{\mbox{\boldmath $\nabla$}}\bar{\eta}) 
\end{array} & 
-{\mbox{\boldmath $1$}}_2\widehat{\wp}_{-} & 
{\mbox{\boldmath $0$}}_2 & {\mbox{\boldmath $0$}}_2 
\end{array}\right),   	
\end{equation}  
\end{subequations} 
where the spatially dependent refractive index 
$n\left( {\mbox{\boldmath $r$}}_\perp, z \right)$ fluctuates around the constant, 
or mean, value $n_0$, and  
\begin{equation}
\widehat{\wp}_x = -i\bar{\lambda} \frac{\partial}{\partial x}, \quad 
\widehat{\wp}_y = -i\bar{\lambda} \frac{\partial}{\partial y}, \quad 
\widehat{\wp}_{+} = \widehat{\wp}_x + i\widehat{\wp}_y, \quad 
\widehat{\wp}_{-} = \widehat{\wp}_x - i\widehat{\wp}_y.    
\end{equation} 
Note that 
\begin{subequations} 
\begin{equation}  
\mathcal{B}\widehat{\mathcal{E}} = \widehat{\mathcal{E}}\mathcal{B}, 
\label{BEBO-a}
\end{equation} 
\begin{equation} 
\mathcal{B}\widehat{\mathcal{O}} = -\widehat{\mathcal{O}}\mathcal{B}.  
\label{BEBO-b}
\end{equation} 
\end{subequations} 
It is seen that the Maxwell vector wave optical Hamiltonian $\widehat{\mathcal{H}}$  in Eq. (\ref{beamoptH}) has the structure of the Dirac Hamiltonian for the electron.  The mean refractive index, $n_0$, has the coefficient matrix $\mathcal{B}$ in $\widehat{\mathcal{H}}$ analogous to the electron mass $m$ having the beta matrix as the coefficient in the Dirac Hamiltonian.  The origin of this result can be traced to making $\mathcal{B}$ the coefficient of $\partial_z$ in Eq. (\ref{phirepeqn}) through the transformation in Eq. (\ref{phitransf}).  Now, with $\widehat{\mathcal{H}}$ having the structure of the Dirac Hamiltonian, it is possible to study the propagation of the electromagnetic beam along the $z$-axis by expanding $\widehat{\mathcal{H}}$ as a series in the parameter $1/n_0$ corresponding to paraxial and higher order approximations adopting the Foldy-Wouthuysen transformation (FWT) technique.  In the relativistic quantum mechanics the FWT technique is applied to expand the Dirac Hamiltonian for the electron as a series in $1/m$ corresponding to nonrelativistic and higher order approximations (Foldy and Wouthuysen, 1950; see also Bjorken and Drell, 1964).  The FWT technique is a successive diagonalization procedure originally developed for the Dirac electron theory.  It can be used to handle matrix evolution equations which can be rewritten in the form of the Dirac equation.  Examples are the Feshbach-Villars form of the Klein-Gordon equation (see Bjorken and Drell, 1964), Helmholtz equation (see, 
{\em e.g.}, Jagannathan and Khan, 2019), and the matrix representation of the Maxwell equations (Khan, 2005; see also Appendix C in this article).  A fairly detailed account of the FWT technique in optics can be found in (Khan, 2008).  

As an application of Eqs. (\ref{beamoptH}) we shall derive in Appendix C the MSS matrix substitution rule for transition from the Helmholtz scalar wave optics to the Maxwell vector wave optics.  This rule was derived originally in (Mukunda et al., 1983) using the relativistic symmetry of the Maxwell equations.  It has been rederived in (Khan, 2016b) using a four dimensional approximation of the eight dimensional representation of the Maxwell equations in Eq. (\ref{KhanRep}), essentially keeping only the four dimensional upper diagonal block along the main diagonal.  The derivation given here is not based on any approximation.  The rule says that whenever, an evolution operator $\mathcal{G}\left({\mbox{\boldmath $r$}}_\perp,\widehat{\wp}_\perp\right)$ is used to describe the propagation of a monochromatic paraxial beam through any ideal linear optical system in scalar wave optics, {\em i.e.},   
\begin{equation} 
\psi_{{\rm out}} 
= \mathcal{G}\left({\mbox{\boldmath $r$}}_\perp,\widehat{\wp}_\perp\right) \psi_{{\rm in}}, 
\end{equation} 
the same system can be described completely in vector wave optics by a matrix function of $({\mbox{\boldmath $r$}}_\perp,\widehat{\wp}_\perp)$ obtained from 
$\mathcal{G}\left({\mbox{\boldmath $r$}}_\perp,\widehat{\wp}_\perp\right)$ by the simple substitution 
\begin{equation} 
{\mbox{\boldmath $r$}}_\perp \longrightarrow {\mbox{\boldmath $Q$}}_\perp 
= {\mbox{\boldmath $r$}}_\perp 
+ \frac{\bar{\lambda}}{n_c}{\mbox{\boldmath $G$}}_\perp\,,   
\end{equation}    
where $\lambda = 2\pi\bar{\lambda}$ is the wavelength of the monochromatic beam in vacuum and $n_c$ is the constant refractive index of the homogeneous medium in which the beam travels before and after passing through the system.  The MSS substitution rule enables us to obtain the vector wave optics readily from the solutions of the scalar wave equation instead of having to face the more difficult problem of solving the Maxwell equations.  It plays a central role in understanding the Fourier optics for the Maxwell field, Gaussian Maxwell beams, cross polarization in laser beams, etc., (Mukunda et al., 1985a; Simon et al., 1986, 1987; Khan, 2023a,b, 2024) 

\section{Conclusion} 
  
We deduced a new eight dimensional matrix representation of the Maxwell equations for a homogeneous medium and generalized it to the case of a linear inhomogeneous medium. This new representation is equivalent, by a unitary transformation, to the eight dimensional matrix representation obtained earlier by one of us (Khan, 2005), heuristically, by inspection of the equations obeyed by the Riemann-Silberstein-Weber (RSW) vector.  We have started with the matrix representation provided naturally by the linear Maxwell's equations themselves, but underdetermined, filling up the gaps in the matrix by symmetry considerations following the work of (Bocker and Frieden, 1993, 2018).  Derivations of the inhomogeneous wave equations for the electric and magnetic fields and the charge continuity equation from the new matrix representation validate our choice of the undetermined matrix elements.  Using a unitary transformation, based on the representation theory of the Pauli algebra, we have reduced the eight dimensional matrix representation of the Maxwell equations for a homogeneous medium to a direct sum of four Pauli matrix blocks.  This process of reduction of the representation to a direct sum of Pauli matrix blocks leads automatically to the definition of the basis vector $\Psi$ of the representation in terms of the RSW vector.  The RSW vector has been in use for a long time as the starting point for obtaining six dimensional matrix representations of the Maxwell equations, mostly, for the vacuum.  In this sense, the RSW vector is a natural choice to to start with for building matrix representations of the Maxwell equations.  The structure of the time evolution equation resulting from this matrix representation for an inhomogeneous medium is analogous to the time evolution equation in quantum  mechanics for a system with an unperturbed Hamiltonian and a perturbation Hamiltonian.  This suggests that electromagnetic wave propagation in an inhomogeneous medium could be studied using techniques borrowed from quantum mechanics.  In the case of a homogeneous medium, which corresponds to the unperturbed Hamiltonian, the new representation reduces to a block diagonal form with four Pauli matrix blocks along the main diagonal, with the other off-diagonal blocks being null matrices.  Pauli matrices have a simple algebra.  This aspect of the new representation should be very useful in applications.  The Maxwell vector wave optical Hamiltonian corresponding to the propagation of a monochromatic beam in an inhomogeneous medium has the structure of the Dirac Hamiltonian for the electron.  Hence, it is possible to study the propagation of the beam using the paraxial and higher order approximations adopting the Foldy-Wouthuysen transformation (FWT) technique.  In relativistic quantum mechanics the FWT technique is used to expand the Dirac Hamiltonian for the electron in a series of nonrelativistic and higher order approximations.  As an application of the new representation we have derived the Mukunda-Simon-Sudarshan (MSS) matrix substitution rule (Mukunda et al., 1983) for transition from the Helmholtz scalar wave optics to the Maxwell vector wave optics.  The MSS rule was obtained originally by analysing the relativistic symmetry of the Maxwell equations.  An earlier derivation of the MSS substitution rule 
(Khan, 2016b) is based on a four dimensional approximation of the eight dimensional representation in (Khan, 2005).  Further applications of the proposed matrix  representation of the Maxwell equations will be presented elsewhere.  

\bigskip 

\bigskip 

\noindent 
{\bf Declaration of competing interest:} 
The authors declare that they have no known competing financial interests or personal relationships that could have appeared to influence the work reported in this paper.

\medskip  

\noindent
{\bf Acknowledgement:}
We are very much thankful to a reviewer whose extensive comments and many constructive suggestions have vastly improved the presentation of our paper.  

\bigskip  

\bigskip 

\noindent 
{\bf Appendix A: Direct sum and direct product of matrices}  \\

\noindent 
If $A$ and $B$ are two square matrices of dimensions $m$ and $n$, respectively, then the direct sum of $A$ and $B$, written as $A \oplus B$, is an $(m+n) \times (m+n)$ matrix given by 
\begin{equation}
A \oplus B 
= \left(\begin{array}{cc} 
A & \mbox{\boldmath $0$}_{m\times n} \\ 
\mbox{\boldmath $0$}_{n\times m} & B 
\end{array}\right),   
\end{equation} 
where $\mbox{\boldmath $0$}_{m\times n}$ and $\mbox{\boldmath $0$}_{n\times m}$ are, 
respectively, $m\times n$ and $n\times m$ null matrices.  The definition can be extended to several matrices:  If $A_1$, $A_2$, $A_3$, $\ldots$, $A_{k-1}$, $A_k$ are square matrices of dimensions $n_1$, $n_2$, $n_3$, $\ldots$, $n_{k-1}$, $n_k$, respectively, then, their direct sum, 
$A_1 \oplus A_2 \oplus A_3 \cdots A_{k-1} \oplus A_k$ is the square matrix of dimension $\left(n_1 + n_2 + n_3 + \cdots + n_{k-1} + n_k\right)$ with $A_1$, $A_2$, 
$A_3$, $\ldots$, $A_{k-1}$, $A_k$, along the main diagonal and all other entries being zero.  The definition can be extended in a straightforward way if some, or all, of the matrices in the sum are rectangular. \\ 

The direct product of two matrices is defined as follows.  In general, if $A$ is 
$k \times \ell$ matrix and $B$ is an $m \times n$ matrix, then the direct product 
of $A$ and $B$, written as $A \otimes B$, is a $km \times \ell n$ matrix given by  
\begin{equation}
A \otimes B 
= \left(\begin{array}{cccc}  
A_{11} B & A_{12} B & \cdots & A_{1\ell} B \\ 
A_{21} B & A_{22} B & \cdots & A_{2\ell} B \\ 
\cdot    & \cdot    & \cdots & \cdot    \\ 
\cdot    & \cdot    & \cdots & \cdot    \\
\cdot    & \cdot    & \cdots & \cdot    \\ 
A_{k1} B & A_{n2} B & \cdots & A_{k\ell} B 
\end{array}\right).   
\label{A1} 
\end{equation}  
If $A$ and $B$ are $m \times m$ and $n \times n$ square matrices, respectively, then 
$A \otimes B$ is an $mn \times mn$ square matrix.  Note that 
$(A \otimes B)^{\dagger} = A^{\dagger} \otimes B^{\dagger}$, and 
$(A \otimes B)(C \otimes D) = (AC) \otimes (BD)$ when the products $AC$ and $BD$
exist.  If $A$ and $C$ are square matrices of dimension $m$ and $B$ and $D$ are square matrices of dimension $n$, then, $(A \otimes B)(C \otimes D) 
= (AC) \otimes (BD)$ exists and is an $mn \times mn$ square matrix.    

\bigskip 

\noindent 
{\bf Appendix B: Derivation of the generalized Helmholtz equations} 

\medskip 

\noindent 
As seen in Section 3, for a time-independent source-free inhomogeneous medium we have 
\begin{equation} 
\frac{\partial^2{\mathcal{F}}}{\partial t^2} 
= \left(\begin{array}{cc}
{\mathcal{M}}_{12}{\mathcal{M}}_{21} & {\mbox{\boldmath $0$}} \\ 
{\mbox{\boldmath $0$}} & {\mathcal{M}}_{21}{\mathcal{M}}_{12}   
\end{array} \right){\mathcal{F}}, 
\label{B-tiNCd2Fdt2}     
\end{equation}   
where 
\begin{eqnarray}
{\mathcal{M}}_{12} 
& = & v\left(\begin{array}{cccc} 
0 & -\partial_z + \partial_z\bar{\mu} 
& \partial_y - \partial_y\bar{\mu} & -\partial_x - \partial_x\bar{\mu} \\ 
\partial_z - \partial_z\bar{\mu} & 0 
& -\partial_x + \partial_x\bar{\mu} & -\partial_y - \partial_y\bar{\mu} \\ 
-\partial_y + \partial_y\bar{\mu} &  \partial_x - \partial_x\bar{\mu} 
& 0 & -\partial_z - \partial_z\bar{\mu} \\ 
\partial_x + \partial_x\bar{\mu} &  \partial_y + \partial_y\bar{\mu} 
& \partial_z + \partial_z\bar{\mu} & 0 
\end{array} \right),  
\nonumber \\  
{\mathcal{M}}_{21} 
& = & v\left(\begin{array}{cccc}
0 & \partial_z - \partial_z\bar{\epsilon}  
& -\partial_y + \partial_y\bar{\epsilon} & \partial_x + \partial_x\bar{\epsilon} \\ 
-\partial_z + \partial_z\bar{\epsilon} & 0 
& \partial_x - \partial_x\bar{\epsilon} & \partial_y + \partial_y\bar{\epsilon} \\ 
\partial_y - \partial_y\bar{\epsilon} & -\partial_x + \partial_x\bar{\epsilon}
& 0 & \partial_z + \partial_z\bar{\epsilon} \\ 
-\partial_x - \partial_x\bar{\epsilon} & -\partial_y - \partial_y\bar{\epsilon} 
& -\partial_z - \partial_z\bar{\epsilon} & 0 
\end{array} \right).  
\label{B-NCdFdt}
\end{eqnarray}  	
For the first component of $\mathcal{F}$, namely $\sqrt{\epsilon}E_x$ apart 
from the constant factor $1/\sqrt{2}$, we get from Eqs. (\ref{B-tiNCd2Fdt2}) and  
(\ref{B-NCdFdt})   
\begin{eqnarray} 
\frac{\partial^2\left(\sqrt{\epsilon}E_x\right)}{\partial t^2} 
& = & v\left\{\left(-\partial_z + \partial_z\bar{\mu}\right)
v\left(-\partial_z + \partial_z\bar{\epsilon}\right)\right. \nonumber \\ 
&   & \qquad + \left(\partial_y - \partial_y\bar{\mu}\right)
v\left(\partial_y - \partial_y\bar{\epsilon}\right)  \nonumber \\ 
&   & \qquad \left. + \left(-\partial_x - \partial_x\bar{\mu}\right)
v\left(-\partial_x - \partial_x\bar{\epsilon}\right)\right\} 
\left(\sqrt{\epsilon}E_x \right)  \nonumber \\ 
&   & \qquad + v\left\{\left(\partial_y - \partial_y\bar{\mu}\right)
v\left(-\partial_x + \partial_x\bar{\epsilon}\right)\right.  \nonumber \\
&   & \qquad \left. + \left(-\partial_x - \partial_x\bar{\mu}\right)
v\left(-\partial_y - \partial_y\bar{\epsilon}\right)\right\} 
\left(\sqrt{\epsilon}E_y\right)   \nonumber \\
&   & \qquad + v\left\{\left(-\partial_z + \partial_z\bar{\mu}\right)
v\left(\partial_x - \partial_x\bar{\epsilon}\right)\right.  \nonumber \\ 
&   & \qquad \left. + \left(-\partial_x - \partial_x\bar{\mu}\right)
v\left(-\partial_z - \partial_z\bar{\epsilon}\right)\right\} 
\left(\sqrt{\epsilon}E_z\right).  
\end{eqnarray}  
Simplifying this equation leads us finally to  
\begin{eqnarray}
\frac{1}{v^2}\frac{\partial^2 E_x}{\partial t^2} 
& = & {\mbox{\boldmath $\nabla$}}^2 E_x 
+ 2\left\{\left(\partial^2_{xx}\bar{\epsilon}\right)E_x 
+ \left(\partial^2_{xy}\bar{\epsilon}\right)E_y 
+ \left(\partial^2_{xz}\bar{\epsilon}\right)E_z \right. \nonumber \\
&   & \qquad + \left. \left(\partial_x\bar{\epsilon}\right)\partial_xE_x 
+ \left(\partial_y\bar{\epsilon}\right)\partial_xE_y  
+ \left(\partial_z\bar{\epsilon}\right)\partial_xE_z\right\}  \nonumber \\
&   & \qquad + \left\{\partial_y\bar{\mu}
\left(\partial_xE_y - \partial_yE_x\right) 
-\partial_z\bar{\mu}\left(\partial_zE_x - \partial_xE_z \right)\right\}  
\nonumber \\ 
& = & {\mbox{\boldmath $\nabla$}}^2 E_x 
+ 2\frac{\partial}{\partial x}\left({\mbox{\boldmath $\nabla$}}
\bar{\epsilon}\cdot{\mbox{\boldmath $E$}}\right) 
+ 2\left({\mbox{\boldmath $\nabla$}}\bar{\mu}\times
({\mbox{\boldmath $\nabla$}}\times{\mbox{\boldmath $E$}})\right)_x,   
\label{ExEq}
\end{eqnarray} 
where 
\begin{eqnarray}
\partial^2_{xx} 
& = & \frac{\partial^2}{\partial x^2},  \quad    
\partial^2_{xy} 
= \frac{\partial^2}{\partial x\partial y} 
= \frac{\partial^2}{\partial y\partial x} 
= \partial^2_{yx},  
\nonumber \\  
\partial^2_{xz} 
& = & \frac{\partial^2}{\partial x\partial z} 
= \frac{\partial^2}{\partial z\partial x} 
= \partial^2_{zx}. 
\end{eqnarray} 
Working out the equations for the second and third components of $\mathcal{F}$ from Eqs. (\ref{B-tiNCd2Fdt2}) and (\ref{B-NCdFdt}) we obtain equations for $E_y$ and $E_z$ similar to Eq. (\ref{ExEq}) with $x$ replaced by $y$ and $z$, respectively.  Thus, it is found that the resulting three equations are the three components of the equation 
\begin{equation} 
{\mbox{\boldmath $\nabla$}}^2{\mbox{\boldmath $E$}} 
- \epsilon\mu\frac{\partial^2{\mbox{\boldmath $E$}}}{\partial t^2} 
+ ({\mbox{\boldmath $\nabla$}}\ln{\mu})\times({\mbox{\boldmath $\nabla$}}
\times{\mbox{\boldmath $E$}}) 
+ {\mbox{\boldmath $\nabla$}}(({\mbox{\boldmath $\nabla$}}
\ln{\epsilon})\cdot{\mbox{\boldmath $E$}}) = 0.   
\label{Helmholtz-BW}
\end{equation} 
which is the generalized Helmholtz equation given in (Born and Wolf, 1999).  Using the well known vector calculus identities 
\begin{eqnarray} 
&   & {\mbox{\boldmath $\nabla$}}({\mbox{\boldmath $A$}}\cdot{\mbox{\boldmath $B$}}) 
= ({\mbox{\boldmath $A$}}\cdot{\mbox{\boldmath $\nabla$}}){\mbox{\boldmath $B$}} 
+ ({\mbox{\boldmath $B$}}\cdot
{\mbox{\boldmath $\nabla$}}){\mbox{\boldmath $A$}}  
\nonumber \\ 
&   &  \qquad \qquad \qquad \qquad 
+ {\mbox{\boldmath $A$}}\times({\mbox{\boldmath $\nabla$}} 
\times {\mbox{\boldmath $B$}}) 
+ {\mbox{\boldmath $B$}}\times({\mbox{\boldmath $\nabla$}}
\times{\mbox{\boldmath $A$}}),  
\nonumber \\ 
&   & {\mbox{\boldmath $\nabla$}}\times({\mbox{\boldmath $\nabla$}}\varphi) = 0,   
\end{eqnarray} 
where $\mbox{\boldmath $A$}$ and $\mbox{\boldmath $B$}$ are vector functions of $\mbox{\boldmath $r$}$, and $\varphi$ is a scalar function of 
$\mbox{\boldmath $r$}$, we can write Eq. (\ref{Helmholtz-BW}) as 
\begin{eqnarray} 
&   & {\mbox{\boldmath $\nabla$}}^2{\mbox{\boldmath $E$}} 
- \epsilon\mu\frac{\partial^2{\mbox{\boldmath $E$}}}{\partial t^2}  
+ (({\mbox{\boldmath $\nabla$}}\ln{\epsilon})\cdot
{\mbox{\boldmath $\nabla$}}){\mbox{\boldmath $E$}} 
+ ({\mbox{\boldmath $E$}}\cdot{\mbox{\boldmath $\nabla$}})
({\mbox{\boldmath $\nabla$}}\ln{\epsilon})  \nonumber \\ 
&   & \qquad \qquad \qquad \qquad 
+ ({\mbox{\boldmath $\nabla$}}(\ln{\epsilon}+\ln{\mu}))\times
({\mbox{\boldmath $\nabla$}}\times{\mbox{\boldmath $E$}}) = 0,    
\label{Helmholtz-M}
\end{eqnarray} 
the form of the generalized Helmholtz equation derived in 
(Mazharimousavi et al., 2013).  It is to be noted that the equation for the fourth component, or the null component, of $\mathcal{F}$ leads to $0 = 0$ exactly.  The equations for the fifth to seventh components of $\mathcal{F}$ lead to the generalized Helmholtz equation for the magnetic field given in 
(Born and Wolf, 1999).  The equation for the eighth, or the null component, of $\mathcal{F}$, leads to $0 = 0$ exactly.  

\bigskip 

\noindent 
{\bf Appendix C: Mukunda-Simon-Sudarshan substitution rule} 

\medskip  

\noindent 
In Section 4 we have seen that for a monochromatic electromagnetic beam moving along the $z$-axis in an inhomogeneous medium with a spatially dependent refractive index $n\left({\mbox{\boldmath $r$}}_\perp, z \right)$ the $z$-evolution equation is given by  
\begin{eqnarray} 
i\bar{\lambda} \frac{\partial}{\partial z} 
\left(\begin{array}{c} 
\bar{\phi}^{+}({\mbox{\boldmath $r$}}_\perp,z) \\
\bar{\phi}^{-}({\mbox{\boldmath $r$}}_\perp,z)   
\end{array}\right) 
& = & \widehat{\mathcal{H}}
\left(\begin{array}{c} 
\bar{\phi}^{+}({\mbox{\boldmath $r$}}_\perp,z) \\
\bar{\phi}^{-}({\mbox{\boldmath $r$}}_\perp,z)   
\end{array}\right),  
\nonumber \\  
\widehat{\mathcal{H}} 
& = & -n_0\mathcal{B} + \widehat{\mathcal{E}} + \widehat{\mathcal{O}},  	
\label{C-beamoptH}
\end{eqnarray} 
where $n_0$ is the constant mean value of the refractive index around which it fluctuates, and $\widehat{\mathcal{E}}$ and $\widehat{\mathcal{O}}$ are as given 
in Eq. (\ref{beamoptH}).  Now, let us write 
\begin{equation}
\left(\begin{array}{c} 
\bar{\phi}^{+}({\mbox{\boldmath $r$}}_\perp,z) \\
\bar{\phi}^{-}({\mbox{\boldmath $r$}}_\perp,z)   
\end{array}\right)^{(1)} 
= \exp{\left(-\frac{\mathcal{B}\widehat{\mathcal{O}}}{2n_0}\right)} 
\left(\begin{array}{c} 
\bar{\phi}^{+}({\mbox{\boldmath $r$}}_\perp,z) \\
\bar{\phi}^{-}({\mbox{\boldmath $r$}}_\perp,z)   
\end{array}\right).    
\label{FWT}
\end{equation}  
This is the first of a series of FW transformations.  The result of this transformation is, after quite some algebra,  
\begin{eqnarray} 
i\bar{\lambda} \frac{\partial}{\partial z}
\left(\begin{array}{c} 
\bar{\phi}^{+}({\mbox{\boldmath $r$}}_\perp,z) \\
\bar{\phi}^{-}({\mbox{\boldmath $r$}}_\perp,z)   
\end{array}\right)^{(1)}  
& = & \widehat{\mathsf{H}}  
\left(\begin{array}{c} 
\bar{\phi}^{+}({\mbox{\boldmath $r$}}_\perp,z) \\
\bar{\phi}^{-}({\mbox{\boldmath $r$}}_\perp,z)   
\end{array}\right)^{(1)},  
\nonumber \\ 
\widehat{\mathsf{H}} 
& \approx & -n\left( {\mbox{\boldmath $r$}}_\perp, z \right) 
+ \frac{\widehat{\wp}_\perp^2}{2n_0}, 
\label{scalarH} 
\end{eqnarray} 
where we have dropped all the matrix terms and kept only the leading order terms to obtain the scalar wave optical Hamiltonian $\widehat{\mathsf{H}}$.  This is the scalar paraxial approximation.  If one uses further FW transformations and drops the matrix terms one gets higher order terms in the scalar wave optical Hamiltonian corresponding to aberrations.  

We shall now consider a monochromatic paraxial beam propagating through an ideal linear optical system along its optic axis in the forward $z$-direction, or in other words, $+z$-direction.  Let the scalar wave optical Hamiltonian of the system be 
$\widehat{\mathsf{H}}\left({\mbox{\boldmath $r$}}_\perp,
\widehat{{\mbox{\boldmath $\wp$}}}_\perp,z\right)$ as in Eq. (\ref{scalarH}).  Since 
$\widehat{\wp}_\perp/2n_0 \ll 1$, $\exp{\left(-\mathcal{B}\widehat{\mathcal{O}}/2n_0\right)} \approx 1$.  Then, from Eqs. (\ref{FWT}) and (\ref{scalarH}), using the Magnus exponential solution of differential equation for a linear operator (see, {\em e.g.}, Blanes et al., 2009; Jagannathan and Khan, 1996, 2019), we have  
\begin{eqnarray} 
\left(\begin{array}{c} 
\bar{\phi}^{+}({\mbox{\boldmath $r$}}_\perp,z^{\prime\prime}) \\ 
\bar{\phi}^{-}({\mbox{\boldmath $r$}}_\perp,z^{\prime\prime})    
\end{array}\right)  
& = & \mathbb{P}\left[ \exp\left\{-\frac{i}{\bar{\lambda}} 
\int_{z^\prime}^{z^{\prime\prime}} dz\,\widehat{\mathsf{H}} 
\left({\mbox{\boldmath $r$}}_\perp,\widehat{{\mbox{\boldmath $\wp$}}}_\perp, z\right)\right\} \right] 
\left(\begin{array}{c} 
\bar{\phi}^{+}({\mbox{\boldmath $r$}}_\perp,z^\prime) \\ 
\bar{\phi}^{-}({\mbox{\boldmath $r$}}_\perp,z^\prime)    
\end{array}\right)  \nonumber \\ 
& = & \exp\left\{ -\frac{i}{\bar{\lambda}}\xi\left({\mbox{\boldmath $r$}}_\perp,
\widehat{{\mbox{\boldmath $\wp$}}}_\perp; z^{\prime\prime},z^\prime\right) \right\} 
\left(\begin{array}{c} 
\bar{\phi}^{+}({\mbox{\boldmath $r$}}_\perp,z^\prime) \\
\bar{\phi}^{-}({\mbox{\boldmath $r$}}_\perp,z^\prime)    
\end{array}\right)  \nonumber \\ 
& = & \mathcal{G}\left({\mbox{\boldmath $r$}}_\perp,
\widehat{{\mbox{\boldmath $\wp$}}}_\perp; z^{\prime\prime},z^\prime\right)
\left(\begin{array}{c} 
\bar{\phi}^{+}({\mbox{\boldmath $r$}}_\perp,z^\prime) \\
\bar{\phi}^{-}({\mbox{\boldmath $r$}}_\perp,z^\prime)    
\end{array}\right)  \nonumber \\   
\end{eqnarray}  
where $\mathbb{P}$ stands for $z$-ordering.  Since  
$\bar{\phi}^{+}({\mbox{\boldmath $r$}}_\perp,z)$ and 
$\bar{\phi}^{-}({\mbox{\boldmath $r$}}_\perp,z)$ are linear combinations of the components of $\sqrt{\epsilon}\,{\mbox{\boldmath $E$}}$ and 
${\mbox{\boldmath $B$}}/\sqrt{\mu}$ with numerical coefficients we can write 
\begin{equation}
\left(\begin{array}{c} 
\sqrt{\epsilon}\,{\mbox{\boldmath $E$}}\left( z^{\prime\prime} \right) \\ 
\frac{1}{\sqrt{\mu}}\,{\mbox{\boldmath $B$}}\left( z^{\prime\prime} \right)  
\end{array}\right) 
= \mathcal{G}\left({\mbox{\boldmath $r$}}_\perp,
\widehat{{\mbox{\boldmath $\wp$}}}_\perp; z^{\prime\prime},z^\prime\right) 
\left(\begin{array}{c} 
\sqrt{\epsilon}\,{\mbox{\boldmath $E$}}\left( z^{\prime} \right) \\ 
\frac{1}{\sqrt{\mu}}\,{\mbox{\boldmath $B$}}\left( z^{\prime} \right)  
\end{array} \right),     
\label{scalar} 
\end{equation} 
as in the scalar wave optics.   

Let us now consider a homogeneous medium with a constant refractive index $n_c$.  
From Eq. (\ref{beamoptH}) we have   
\begin{subequations} 
\label{beamoptH0} 
\begin{equation} 
i\bar{\lambda} \frac{\partial}{\partial z} 
\left(\begin{array}{c} 
\bar{\phi}^{+}({\mbox{\boldmath $r$}}_\perp,z) \\
\bar{\phi}^{-}({\mbox{\boldmath $r$}}_\perp,z)   
\end{array}\right) 
= \widehat{\mathcal{H}} 
\left(\begin{array}{c} 
\bar{\phi}^{+}({\mbox{\boldmath $r$}}_\perp,z) \\
\bar{\phi}^{-}({\mbox{\boldmath $r$}}_\perp,z)   
\end{array}\right),  
\end{equation} 
\begin{equation}  
\widehat{\mathcal{H}} 
= -n_c\mathcal{B} + \widehat{\mathcal{O}},  	
\end{equation} 
\begin{equation} 
\widehat{\mathcal{O}} 
= \left(\begin{array}{cccc}  
{\mbox{\boldmath $0$}} & {\mbox{\boldmath $0$}} & 
{\mbox{\boldmath $1$}} \widehat{\wp}_{-} & {\mbox{\boldmath $0$}} \\ 
{\mbox{\boldmath $0$}} & {\mbox{\boldmath $0$}} &	
{\mbox{\boldmath $0$}} & {\mbox{\boldmath $1$}} \widehat{\wp}_{+}  \\ 
-{\mbox{\boldmath $1$}} \widehat{\wp}_{+} & {\mbox{\boldmath $0$}} & 
{\mbox{\boldmath $0$}} & {\mbox{\boldmath $0$}} \\ 
{\mbox{\boldmath $0$}} & -{\mbox{\boldmath $1$}} \widehat{\wp}_{-} & 
{\mbox{\boldmath $0$}} & {\mbox{\boldmath $0$}} 
\end{array} \right).   
\end{equation}  
\end{subequations} 
With $\widehat{\wp}_z^2 = \left(n_c^2 - \widehat{\wp}_\perp^2\right)$, we have 
\begin{equation}
\widehat{\mathcal{H}}^2 = {\mbox{\boldmath $1$}}_8\widehat{\wp}_z^2.      
\label{Hsquare} 
\end{equation} 
Define 
\begin{equation} 
\mathfrak{T} 
= \frac{\left({\mbox{\boldmath $1$}}_8\widehat{\wp}_z - \mathcal{B}\widehat{\mathcal{H}}\right)}
{\sqrt{2\widehat{\wp}_z\left(n_c + \widehat{\wp}_z\right)}}. 
\label{defmathfrakT} 
\end{equation} 
Then, 
\begin{equation} 
{\mathfrak{T}}^{-1} 
= \frac{\left({\mbox{\boldmath $1$}}_8\widehat{\wp}_z
-\widehat{\mathcal{H}}\mathcal{B}\right)}
{\sqrt{2\widehat{\wp}_z \left( n_c + \widehat{\wp}_z \right)}}.  
\end{equation} 
To see this, first note that $\widehat{\wp}_z$ commutes with $\mathcal{B}$ and $\widehat{\mathcal{H}}$, and $\mathcal{B}^2 = \mbox{\boldmath $1$}_8$.  Then, using Eqs. (\ref{BEBO-b}) and (\ref{Hsquare}), observe 
\begin{eqnarray} 
\left({\mbox{\boldmath $1$}}_8\widehat{\wp}_z 
-\mathcal{B}\widehat{\mathcal{H}}\right)
\left({\mbox{\boldmath $1$}}_8\widehat{\wp}_z
-\widehat{\mathcal{H}}\mathcal{B}\right) 
& = & {\mbox{\boldmath $1$}}_8\widehat{\wp}_z^2 
-\widehat{\wp}_z\widehat{\mathcal{H}}\mathcal{B} 
-\mathcal{B}\widehat{\mathcal{H}}\widehat{\wp}_z 
+\mathcal{B}\widehat{\mathcal{H}}^2\mathcal{B} 
\nonumber \\ 
& = & \left({\mbox{\boldmath $1$}}_8\widehat{\wp}_z^2 
+ \mathcal{B}\widehat{\mathcal{H}}^2\mathcal{B}\right)
- \widehat{\wp}_z(\widehat{\mathcal{H}}\mathcal{B} 
+ \mathcal{B}\widehat{\mathcal{H}})  
\nonumber \\ 
& = & 2{\mbox{\boldmath $1$}}_8\widehat{\wp}_z(n_c + \widehat{\wp}_z).  
\end{eqnarray}
Now, we have  
\begin{equation}
\mathfrak{T}\widehat{\mathcal{H}}{\mathfrak{T}}^{-1} 
= -\mathcal{B}\widehat{\wp}_z.  
\label{diagmathcalH}
\end{equation} 
This follows from the identity 
\begin{equation}
\left({\mbox{\boldmath $1$}}_8\widehat{\wp}_z - \mathcal{B}\widehat{\mathcal{H}}\right)\widehat{\mathcal{H}} 
= -\mathcal{B}\widehat{\wp}_z\left({\mbox{\boldmath $1$}}_8\widehat{\wp}_z - \mathcal{B}\widehat{\mathcal{H}}\right).  
\end{equation} 
We shall take, in the paraxial approximation,   
\begin{equation} 
\widehat{\wp}_z 
= +\sqrt{n_c^2 - \widehat{\wp}_\perp^2}\,\approx 
n_c \left( 1 - \frac{\widehat{\wp}_\perp^2}{2n_c^2} \right) 
= n_c - \frac{\widehat{\wp}_\perp^2}{2n_c}. 
\end{equation}  
The square roots in Eq. (\ref{defmathfrakT}) are also to be understood as expansions in terms of $\widehat{\wp}_\perp^2/n_c^2$ to the required degree of accuracy.  Defining 
\begin{eqnarray}
\left(\begin{array}{c} 
\varphi^{+}({\mbox{\boldmath $r$}}_\perp,z) \\
\varphi^{-}({\mbox{\boldmath $r$}}_\perp,z)   
\end{array}\right) 
& = & \mathfrak{T}  
\left(\begin{array}{c} 
\bar{\phi}^{+}({\mbox{\boldmath $r$}}_\perp,z) \\
\bar{\phi}^{-}({\mbox{\boldmath $r$}}_\perp,z)   
\end{array}\right)  
\nonumber \\ 
& = & \frac{1}{2 \sqrt{2\widehat{\wp}_z\left( n_c + \widehat{\wp}_z \right)}} 
\nonumber \\ 
&   & \qquad \times 
\left(\begin{array}{c}
\left(n_c + \widehat{\wp}_z \right) \left( -F_x^{+} + iF_y^{+} \right) 
- \widehat{\wp}_{-} F_z^{+} \\
\left(n_c + \widehat{\wp}_z \right) F_z^{+} 
- \widehat{\wp}_{-} \left(F_x^{+} + i F_y^{+} \right) \\
\left(n_c + \widehat{\wp}_z \right) \left(- F_x^{-} - i F_y^{-} \right) 
- \widehat{\wp}_{+} F_z^{-} \\ 
\left(n_c + \widehat{\wp}_z \right) F_z^{-} 
- \widehat{\wp}_{+} \left(F_x^{-} - i F_y^{-} \right) \\
\left(n_c + \widehat{\wp}_z \right) F_z^{+} 
- \widehat{\wp}_{+} \left(- F_x^{+} + i F_y^{+} \right) \\
\left(n_c + \widehat{\wp}_z \right) \left(F_x^{+} + i F_y^{+} \right) 
- \widehat{\wp}_{+} F_z^{+} \\
\left(n_c + \widehat{\wp}_z \right) F_z^{-} 
- \widehat{\wp}_{-} \left(- F_x^{-} - i F_y^{-} \right) \\
\left(n_c + \widehat{\wp}_z \right) \left(F_x^{-} - i F_y^{-} \right) 
- \widehat{\wp}_{-} F_z^{-}
\end{array} \right),  
\nonumber \\ 
&   &  
\label{varphieqnb}
\end{eqnarray}  
we have, from Eq. (\ref{beamoptH0}) and Eq. (\ref{diagmathcalH}),  
\begin{equation}
i\bar{\lambda} \frac{\partial}{\partial z}
\left(\begin{array}{c} 
\varphi^{+}({\mbox{\boldmath $r$}}_\perp,z) \\
\varphi^{-}({\mbox{\boldmath $r$}}_\perp,z)   
\end{array}\right) 
= -\mathcal{B} \widehat{\wp}_z 
\left(\begin{array}{c} 
\varphi^{+}({\mbox{\boldmath $r$}}_\perp,z) \\
\varphi^{-}({\mbox{\boldmath $r$}}_\perp,z)   
\end{array}\right).    
\label{varphieqna}
\end{equation} 
A plane wave moving close to the positive $z$-direction is represented by 
\begin{eqnarray} 
\left(\begin{array}{c} 
\varphi^{+}({\mbox{\boldmath $r$}}_\perp,z) \\
\varphi^{-}({\mbox{\boldmath $r$}}_\perp,z)   
\end{array}\right) 
& = & \left(\begin{array}{c} 
\varphi_0^{+} \\ \varphi_0^{-}    
\end{array}\right) 
\exp\{ i \left( {\mbox{\boldmath $k$}}_\perp \cdot 
{\mbox{\boldmath $r$}}_\perp + k_zz \right)\},  \nonumber \\ 
&   & \qquad \qquad \qquad k_z > 0, \quad 
\left| k_x \right|, \left| k_y \right| \ll k_z,  
\label{planevarphi}
\end{eqnarray} 
where the components of $\varphi_0^{+}$ and $\varphi_0^{-}$ are constants as seen from Eq. (\ref{planewaveF}) and Eq. (\ref{varphieqnb}).  For a plane wave the relation between the magnitude of the wave vector, $k$, in the medium and $\kappa$, in vacuum, is $k = n_c\kappa$ or $k_\perp^2 + k_z^2 = n_c^2\kappa^2$.  This means that $k^2_z/\kappa^2 = n_c^2 - (k_\perp^2/\kappa^2)$.  The plane wave in  Eq. (\ref{planevarphi}) has to satisfy Eq. (\ref{varphieqna}).  This leads to 
$-k_z/\kappa = -\sqrt{n_c^2 - \left(k_\perp^2/\kappa^2\right)} \approx 
-n_c\left(1 - (k_\perp^2/2k^2)\right)$ for $\varphi^{+}$ and $-k_z/\kappa = \sqrt{n_c^2 - \left(k_\perp^2/\kappa^2\right)} \approx n_c 
\left(1 - (k_\perp^2/2k^2)\right)$ for $\varphi^{-}$.  This implies that, since 
$k_z > 0$ and $k_\perp^2 \ll k^2$, the plane wave moving in the $+z$-direction must be associated with a $\varphi^{-}$ for which all the four components vanish.  This can be easily verified from Eq. (\ref{varphieqnb}) for a plane wave moving in the $+z$-direction and polarized in the $x$-direction.  A paraxial beam with a finite transverse extent in the $xy$-plane propagating predominantly in the forward $z$-direction can be written as a superposition of plane waves of the type in Eq. (\ref{planevarphi}).  Explicitly, for any such beam satisfying the paraxial condition $k_x^2, k_y^2 \ll k^2$ we can write 
\begin{eqnarray}
{\mbox{\boldmath $F$}}^{\pm} 
& = & \exp\{ikz\} \int_{-\infty}^{\infty} \int_{-\infty}^{\infty} dk_x dk_y 
\tilde{\mbox{\boldmath $F$}}^{\pm} \left( k_x,k_y \right)  
\nonumber \\ 
&   & \qquad \qquad \qquad \times \exp\left\{i\left[k_xx + k_yy - 
z\left(\frac{k_x^2 + k_y^2}{2k}\right)\right]\right\}, 
\end{eqnarray} 
or,  
\begin{eqnarray}
\left(\begin{array}{c} 
\varphi^{+} \\ \varphi^{-}   
\end{array}\right) 
& = & \frac{1}{2} \exp\{ikz\} 
\int_{-\infty}^{\infty} \int_{-\infty}^{\infty} dk_x dk_y 
\frac{1}{\sqrt{2{\wp}_z\left(n_c + {\wp}_z\right)}}  \nonumber \\ 
&   & \qquad \qquad \times \left( 
\begin{array}{c}
\left(n_c + {\wp}_z\right) 
\left(-\tilde{F}_x^{+} + i\tilde{F}_y^{+} \right) - {\wp}_{-} \tilde{F}_z^{+} \\
\left(n_c + {\wp}_z\right) \tilde{F}_z^{+} - {\wp}_{-} \left(\tilde{F}_x^{+} 
+ i\tilde{F}_y^{+}\right) \\
\left(n_c + {\wp}_z\right) \left(-\tilde{F}_x^{-} - i\tilde{F}_y^{-}\right) 
- {\wp}_{+} \tilde{F}_z^{-} \\ 
\left(n_c + {\wp}_z\right) \tilde{F}_z^{-} - {\wp}_{+} \left(\tilde{F}_x^{-} 
- i\tilde{F}_y^{-}\right) \\
\left(n_c + {\wp}_z\right) \tilde{F}_z^{+} - {\wp}_{+} \left(-\tilde{F}_x^{+} 
+ i\tilde{F}_y^{+}\right) \\
\left(n_c + {\wp}_z\right) \left(\tilde{F}_x^{+} + i\tilde{F}_y^{+}\right) 
- {\wp}_{+}\tilde{F}_z^{+} \\
\left(n_c + {\wp}_z\right) \tilde{F}_z^{-} - {\wp}_{-} \left(-\tilde{F}_x^{-} 
- i\tilde{F}_y^{-}\right) \\
\left(n_c + {\wp}_z\right) \left(\tilde{F}_x^{-} - i\tilde{F}_y^{-} \right) 
- {\wp}_{-} \tilde{F}_z^{-}
\end{array} \right)  
\nonumber \\ 
&   & \qquad \qquad 
\times\,\exp\left\{i\left[ k_xx + k_yy - z\left(\frac{k_x^2 + k_y^2}{2k} 
\right) \right]\right\},   
\label{Fouriervarphi}
\end{eqnarray}  
with $\wp_{\pm} = n_c \left(k_x \pm ik_y \right)/k$, 
$\wp_z \approx n_c\left(1 - \left(k_\perp^2/2k^2\right)\right)$, and 
$\tilde{\mbox{\boldmath $F$}}^{\pm}\left(k_x,k_y\right)$ being nonzero only within a small range of $k_x,k_y \ll k$.  The vanishing of all the four components of $\varphi^{-}$ for a paraxial beam propagating in the forward $z$-direction in a homogeneous medium implies a set of relationships among the components of the ${\mbox{\boldmath $E$}}$ and ${\mbox{\boldmath $B$}}$ fields of the beam and this leads to interesting consequences.  This relationship can be written compactly as 
\begin{equation}
\left(\begin{array}{c}
\sqrt{\epsilon}\,{\mbox{\boldmath $E$}} \\
\frac{1}{\sqrt{\mu}}\,{\mbox{\boldmath $B$}} 
\end{array}\right) 
= \left(\frac{2 \bar{\lambda}}{\left(n_c + \widehat{\wp}_z\right)} 
\left(G_x{\partial}_x + G_y{\partial}_y\right) + G_z\right) 
\left(\begin{array}{c}
\sqrt{\epsilon}\,{\mbox{\boldmath $E$}} \\
\frac{1}{\sqrt{\mu}}\,{\mbox{\boldmath $B$}} 
\end{array}\right),  
\label{field-conditions}
\end{equation} 
where the matrices $G_x$, $G_y$ and $G_z$ are  
\begin{eqnarray}
G_x & = & \frac{1}{2}
\left(\begin{array}{cc}
-S_2 & S_1 \\ -S_1 & -S_2
\end{array} \right), \qquad 
G_y = \frac{1}{2}
\left(\begin{array}{cc}
S_1 & S_2 \\ -S_2 & S_1
\end{array} \right), 
\nonumber \\ 
S_1 & = & \left(\begin{array}{ccr}
0 & 0 &  0 \\ 0 & 0 & -i \\ 0 & i &  0
\end{array} \right), \qquad
S_2 = \left(\begin{array}{rcc}
0 & 0 & i \\ 0 & 0 & 0 \\ -i & 0 & 0
\end{array} \right), 
\nonumber \\
G_z & = & \left(\begin{array}{crcrcc}
0 &  0 & 0 &  0 & 1 & 0 \\ 
0 &  0 & 0 & -1 & 0 & 0 \\
0 &  0 & 0 &  0 & 0 & 0 \\ 
0 & -1 & 0 &  0 & 0 & 0 \\
1 &  0 & 0 &  0 & 0 & 0 \\ 
0 &  0 & 0 &  0 & 0 & 0 \\
\end{array} \right),  
\end{eqnarray} 
where we are using the same notations for the matrices as in (Simon et al., 1986). 

Let us now consider a monochromatic paraxial beam to be incident from a homogeneous medium of constant refractive index $n_c$ on an ideal linear optical system at the input $xy$-plane at $z = z^\prime$, pass through it, and emerge into that homogeneous medium at the output $xy$-plane at $z = z^{\prime\prime}$.  The beam is moving through the entire system along the optic axis in the forward $z$-direction.  Let the propagation of the beam through the optical system from the input plane to the output plane be described by the scalar wave optics as given by 
Eq. (\ref{scalar}).  Since the output beam at $z = z^{\prime\prime}$ in the homogeneous medium should satisfy  Eq. (\ref{field-conditions}) we can take it to be given by 
\begin{eqnarray}
\left(\begin{array}{c} 
\sqrt{\epsilon}\,{\mbox{\boldmath $E$}} \left(z^{\prime\prime}\right) \\
\frac{1}{\sqrt{\mu}}\,{\mbox{\boldmath $B$}} \left(z^{\prime\prime}\right)  
\end{array}\right) 
& = & \left(\frac{2\bar{\lambda}}{\left(n_c + \widehat{\wp}_z \right)} 
\left(G_x{\partial}_x + G_y{\partial}_y\right) + G_z\right)  
\nonumber \\ 
&   & \qquad \qquad 
\times\,\exp\left\{-\frac{i}{\bar{\lambda}}\xi\left({\mbox{\boldmath $r$}}_\perp,
\widehat{{\mbox{\boldmath $\wp$}}}_\perp; z^{\prime\prime},z^\prime\right)\right\} 
\left(\begin{array}{c}
\sqrt{\epsilon}\,{\mbox{\boldmath $E$}}\left(z^\prime\right) \\
\frac{1}{\sqrt{\mu}}\,{\mbox{\boldmath $B$}}\left(z^\prime\right) 
\end{array}\right)  \nonumber \\ 
& = & \left(\frac{2\bar{\lambda}}{\left(n_c + \widehat{\wp}_z\right)} 
\left(G_x{\partial}_x + G_y{\partial}_y\right) + G_z\right)  
\nonumber \\ 
&   & \qquad \qquad \times\,\mathcal{G}\left({\mbox{\boldmath $r$}}_\perp,
\widehat{{\mbox{\boldmath $\wp$}}}_\perp; z^{\prime\prime},z^\prime\right)
\left(\begin{array}{c} 
\sqrt{\epsilon}\,{\mbox{\boldmath $E$}}\left(z^\prime\right) \\
\frac{1}{\sqrt{\mu}}\,{\mbox{\boldmath $B$}}\left(z^\prime\right) 
\end{array} \right).     
\end{eqnarray} 
After some lengthy straightforward algebra, taking $\widehat{\wp}_z \approx n_c$, 
we arrive at the result 
\begin{eqnarray}
\left(\begin{array}{c} 
\sqrt{\epsilon}\,{\mbox{\boldmath $E$}}\left(z^{\prime\prime}\right) \\
\frac{1}{\sqrt{\mu}}\,{\mbox{\boldmath $B$}}\left(z^{\prime\prime}\right)  
\end{array}\right)  
& = & \exp\left\{\frac{\bar{\lambda}}{n_c}{\mbox{\boldmath $G$}}_\perp \cdot 
{\mbox{\boldmath $\nabla$}}_\perp\right\}
\mathcal{G}\left({\mbox{\boldmath $r$}}_\perp,
\widehat{{\mbox{\boldmath $\wp$}}}_\perp; z^{\prime\prime},z^\prime\right) 
\left(\begin{array}{c}
\sqrt{\epsilon}\,{\mbox{\boldmath $E$}}\left(z^\prime\right) \\
\frac{1}{\sqrt{\mu}}\,{\mbox{\boldmath $B$}}\left(z^\prime\right) 
\end{array}\right)  
\nonumber \\  
& = & \mathcal{G}\left({\mbox{\boldmath $r$}}_\perp 
+ \frac{\bar{\lambda}}{n_c}{\mbox{\boldmath $G$}}_\perp,
\widehat{{\mbox{\boldmath $\wp$}}}_\perp; z^{\prime\prime},z^\prime\right) 
\left(\begin{array}{c}
\sqrt{\epsilon}\,{\mbox{\boldmath $E$}}\left(z^\prime\right) \\
\frac{1}{\sqrt{\mu}}\,{\mbox{\boldmath $B$}}\left(z^\prime\right) 
\end{array}\right)  \nonumber \\ 
& = & \mathcal{G}\left({\mbox{\boldmath $Q$}}_\perp,
\widehat{{\mbox{\boldmath $\wp$}}}_\perp; z^{\prime\prime},z^\prime\right)
\left(\begin{array}{c}
\sqrt{\epsilon}\,{\mbox{\boldmath $E$}}\left(z^\prime\right) \\
\frac{1}{\sqrt{\mu}}\,{\mbox{\boldmath $B$}}\left(z^\prime\right) 
\end{array}\right),  
\nonumber \\   
&   &  \qquad \qquad \qquad \qquad 
\mbox{with}\ {\mbox{\boldmath $Q$}}_\perp 
= {\mbox{\boldmath $r$}}_\perp + \frac{\bar{\lambda}}{n_c} 
{\mbox{\boldmath $G$}}_\perp. 
\end{eqnarray} 
Note that, with $G_x$ and $G_y$ being numerical matrices, 
${\mbox{\boldmath $Q$}}_\perp$ has the  same dimension as 
${\mbox{\boldmath $r$}}_\perp$.  Thus, the vector wave optical $z$-evolution operator is obtained from the scalar wave optical $z$-evolution operator by the substitution 
\begin{equation}  
{\mbox{\boldmath $r$}}_\perp \longrightarrow 
{\mbox{\boldmath $Q$}}_\perp = {\mbox{\boldmath $r$}}_\perp 
+ \frac{\bar{\lambda}}{n_c} {\mbox{\boldmath $G$}}_\perp.  
\end{equation} 
This is the Mukunda-Simon-Sudarshan substitution rule for transition from the Helmholtz scalar wave optics to the Maxwell vector wave optics.  It has been successfully demonstrated for a variety of light beams (Mukunda et al., 1985a; 
Simon et al., 1986, 1987; Khan, 2023a,b; Khan, 2024a).  

\vspace{1cm}  

\section*{References} 
  
\noindent  
Barnett, S.M., 2014.     
Optical Dirac equation.      
New J. Phys. 16, 093008. 
\url{https://doi.org/10.1088/1367-2630/16/9/093008}  \\      

\noindent  
Belkovich, I.V., Kogan, B.L., 2016.         
Utilization of Riemann-Silberstein vectors in electromagnetics.   
Prog. Electromag. Res. B 69, 103-116. 
\url{http://dx.doi.org/10.2528/PIERB16051809}  \\      

\noindent   
Bialynicki-Birula, I., 1994. 
On the wave function of the photon.  
Acta Phys. Pol. A 86, 97–116. 
\url{http://przyrbwn.icm.edu.pl/APP/ABSTR/86/a86-1-8.html}  \\    

\noindent  
Bialynicki-Birula, I., 1996a.     
The photon wave function.     
In: Coherence and Quantum Optics VII. Eds. Eberly, J.H., Mandel, L., Wolf, E.,  pp.313-322 
(Plenum Press). 
\url{http://dx.doi.org/10.1007/978-1-4757-9742-8_38}  \\      

\noindent 
Bialynicki-Birula, I., 1996b.      
Photon wave function.    
In: Progress in Optics XXXVI, Ed. Wolf, E., pp.245-294 
(Elsevier). 
\url{http://dx.doi.org/10.1016/S0079-6638(08)70316-0}  \\   

\noindent 
Bialynicki-Birula, I., Bialynicki-Birula, Z., 2013.           
The role of the Riemann-Silberstein vector in classical and quantum theories of electromagnetism.     
J. Phys. A: Math. Theor. 46, 053001. 
\url{http://dx.doi.org/10.1088/1751-8113/46/5/053001}  \\    

\noindent 
Bjorken, J.D., Drell, S.D., 1964. 
Relativistic Quantum Mechanics. 
(McGraw Hill, New York)  \\ 

\noindent   
Blanes, S., Casas, F., Oteo, J.A., Jos\'{e}, R., 2009.     
The Magnus expansion and some of its applications.  
Phys. Rep. 470, 151-238. 
\url{https://doi.org/10.1016/j.physrep.2008.11.001}  \\  

\noindent 
Bocker, R.P., Frieden, B.R., 1993.      
Solution of the Maxwell field equations in vacuum for arbitrary charge and current 
distributions using the methods of matrix algebra.  
IEEE Trans. Edu. 36, 350-356. 
\url{https://doi.org/10.1109/13.241610}  \\     

\noindent 
Bocker, R.P., Freiden, B.R., 2018.        
A new matrix formulation of the Maxwell and Dirac equations. 
Heliyon. 4, e01033. 
\url{https://doi.org/10.1016/j.heliyon.2018.e01033}  \\       

\noindent 
Bogush, A., Red'kov, V., Tokarevskaya, N., Spix, G., 2009.    
Majorana-Oppenheimer approach to Maxwell electrodynamics in Riemannian space-time.    
arXiv:0905.0261[math-ph]. 
\url{https://doi.org/10.48550/arXiv.0905.0261}  \\    

\noindent 
Born, M., Wolf, E., 1999.    
Principles of Optics.  
(Camb. Univ. Press, 7th Edn.). 
\url{https://doi.org/10.1017/CBO9781139644181}  \\       

\noindent 
Conte, M., Jagannathan, R., Khan, S.A., Pusterla, M., 1996.     
Beam optics of the Dirac particle with anomalous magnetic moment.    
Particle Accelerators. 56, 99-126. 
\url{http://cds.cern.ch/record/307931/files/p99.pdf}  \\       
	
\noindent 
Edmonds, J.D., 1975.      
Comment on the Dirac-like equation for the photon.   
Lett. Nuovo Cim. 13, 185-186. 
\url{https://doi.org/10.1007/BF02742609}  \\ 

\noindent 
Esposito, S., 1998.     
Covariant Majorana formulation of electrodynamics.     
Found. Phys. 28, 231-244. 
\url{http://dx.doi.org/10.1023/A:1018752803368}  \\ 

\noindent 
Foldy, L.L., Wouthuysen, S.A., 1950. 
On the Dirac theory of spin $\frac{1}{2}$ particles and its non-relativistic limit.  Phys. Rev. 78, 29–36. 
\url{https://doi.org/10.1103/PhysRev.78.29}  \\ 

\noindent 
Good, R.H., 1957.        
Particle aspect of the electromagnetic field equations.    
Phys. Rev. 105, 1914-1919. 
\url{https://doi.org/10.1103/PhysRev.105.1914}  \\ 

\noindent 
Hawkes, P.W., 2020.    
Dirac, c and a Supper date.   
Ultramicroscopy 213, 112981. 
\url{https://doi.org/10.1016/j.ultramic.2020.112981}  \\ 
   
\noindent 
Hawkes, P.W., Kasper, E., 2022.      
Principles of Electron Optics: Vol.3 Fundamental Wave Optics.  
(Academic Press, 2nd Edn.). \\  

\noindent  
Jagannathan, R., Simon, R., Sudarshan, E.C.G., Mukunda, N., 1989.     
Quantum theory of magnetic electron lenses based on the Dirac equation.   
Phys. Lett. A 134, 457-464.  
\url{http://dx.doi.org/10.1016/0375-9601(89)90685-3}  \\    

\noindent 
Jagannathan, R., 1990.     
Quantum theory of electron lenses based on the Dirac equation. 
Phys. Rev. A 42, 6674-6689.  
\url{http://dx.doi.org/10.1103/PhysRevA.42.6674}  \\ 

\noindent
Jagannathan, R., Khan, S.A., 1996.   
Quantum theory of the optics of charged particles. 
In: Advances in Imaging and Electron Physics 97, Ed. Hawkes, P.W., pp.257-358    
(Academic Press) 
\url{http://dx.doi.org/10.1016/S1076-5670(08)70096-X}  \\         

\noindent 
Jagannathan, R., 1999.  
The Dirac equation approach to spin-$\frac{1}{2}$ particle beam optics. 
Proc. 15th Advanced ICFA Beam Dynamics Workshop on Quantum Aspects of Beam Physics,  
(Monterey, California, 1998) Ed. Chen, P., pp.670-681. 
(World Scientific, 1999).  
\url{https://arxiv.org/abs/physics/9803042}  \\ 

\noindent 
Jagannathan, R., 2002. 
Quantum mechanics of Dirac particle beam optics: Single-particle theory. 
Proc. 18th Advanced ICFA Beam Dynamics Workshop on Quantum Aspects of Beam Physics.  
(Capri, Italy, 2000) Ed. Chen, P., pp.568-577. 
(World Scientific, 2002)    
\url{https://doi.org/10.1142/9789812777447_0047}  \\ 

\noindent 
Jagannathan, R., 2004.  
Quantum mechanics of Dirac particle beam transport through optical elements with straight and curved optical axes.  
Proc. 28th Advanced ICFA Beam Dynamics and Advanced \& Novel Accelerators Workshop.  
(Hiroshima, Japan, 2003) Eds. Chen, P., Reil, K., pp.13-21. 
(World Scientific, 2004).   
\url{https://doi.org/10.1142/9789812702333_0002}  \\ 

\noindent 
Jagannathan, R., Khan, S.A., 2019.     
Quantum Mechanics of Charged Particle Beam Optics: Understanding Devices from Electron Microscopes to Particle Accelerators.  
(CRC Press)         
\url{https://doi.org/10.1201/9781315232515}  \\   

\noindent  
Jest\"{a}dt, R., Ruggenthaler, M., Oliveira, M.J.T., Rubio, A., Appel, H.,  2019.        
Light-matter interactions within the Ehrenfest-Maxwell-Pauli-Kohn-Sham framework: 
fundamentals, implementation, and nano-optical applications.    
Adv. Phys. 68, 225-333. 
\url{https://doi.org/10.1080/00018732.2019.1695875}  \\   

\noindent 
Jin, S., Liu, N., Ma, C., 2023.  
Quantum simulation of Maxwell's equations via Schr\"{o}dingerisation. 
arXiv:2308.08408[quant.ph].  
\url{https://doi.org/10.48550/arXiv.2308.08408}  \\ 

\noindent 
Khan, S.A., 1997.      
Quantum Theory of Charged-Particle Beam Optics. 
PhD Thesis (University of Madras, Chennai, India).  
(Available from the DSpace of IMSc Library, The Institute of Mathematical 
Sciences, Chennai, where the doctoral research was done) 
\url{http://www.imsc.res.in/xmlui/handle/123456789/75}  \\ 

\noindent 
Khan, S.A., 1999.    
Quantum theory of magnetic quadrupole lenses for spin-$\frac{1}{2}$ particles.
In: Proc. 15th Advanced ICFA Beam Dynamics Workshop on Quantum Aspects of Beam Physics. Monterey, CA, USA, 1998. Ed. Chen, P., pp.682-694.  
(World Scientific, 1999)  
\url{https://arxiv.org/abs/physics/9809032}  \\ 

\noindent 
Khan, S.A., 2002.     
Quantum formalism of beam optics.   
Proc. 18th Advanced ICFA Beam Dynamics Workshop on Quantum Aspects of Beam Physics. 
Capri, Italy, 2000. Ed. Chen, P., pp.517-526. 
(World Scientific, 2002)  
\url{http://dx.doi.org/10.1142/9789812777447_0042}  \\ 

\noindent 
Khan, S.A., 2005.      
An exact matrix representation of Maxwell's equations.   
Phys. Scr. 71, 440-442. 
\url{http://dx.doi.org/10.1238/Physica.Regular.071a00440}  \\       

\noindent 
Khan, S.A., 2008.   
The Foldy-Wouthuysen transformation technique in optics.   
In: Advances in Imaging and Electron Physics 152, Ed. Hawkes, P.W., pp.49-78,     
(Academic Press)  
\url{http://dx.doi.org/10.1016/S1076-5670(08)00602-2}  \\ 
   
\noindent 
Khan, S.A., 2010.     
Maxwell optics of quasiparaxial beams.   
Optik 121, 408-416.     
\url{http://dx.doi.org/10.1016/j.ijleo.2008.07.027}  \\ 

\noindent 
Khan, S.A., 2014.     
Aberrations in Maxwell optics.    
Optik 125, 968-978.  
\url{http://dx.doi.org/10.1016/j.ijleo.2013.07.097}  \\     

\noindent
Khan, S.A., 2016a.
International Year of Light and History of Optics,
in: Advances in Photonics Engineering, Nanophotonics and Biophotonics, Ed. Tanya Scott, pp.1-56. 
(Nova Science Publ.)  \\ 

\noindent 
Khan, S.A., 2016b.   
Passage from scalar to vector optics and the Mukunda-Simon-Sudarshan theory for 
paraxial beams.    
J. Mod. Opt., 1652-1660.  
\url{http://dx.doi.org/10.1080/09500340.2016.1164257}  \\              

\noindent 
Khan, S.A., 2017a.   
Polarization in Maxwell optics. 
Optik 131, 733-748.  
\url{http://dx.doi.org/10.1016/j.ijleo.2016.11.134}  \\      

\noindent
Khan, S.A., 2017b.      
Quantum methodologies in Maxwell optics.    
In: Advances in Imaging an Electron Physics 201, Ed. Hawkes, P.W., pp.57-135.    
(Academic Press) 
\url{http://dx.doi.org/10.1016/bs.aiep.2017.05.003}  \\  

\noindent 
Khan, S.A., 2023a. 
Cross polarization in Gaussian and Bessel light beams. 
Opt. Commun. 545, 129728.   
\url{https://doi.org/10.1016/j.optcom.2023.129728}  \\ 

\noindent 
Khan, S.A., 2023b.  
Anisotropic Airy beams. 
Results in Optics 13, 100569.  
\url{https://doi.org/10.1016/j.rio.2023.100569}  \\ 

\noindent 
Khan, S.A., 2024. 
Cross polarization in anisotropic Gaussian light beams. 
Indian J. Phys. 98, 3699-3705.  
\url{https://doi.org/10.1007/s12648-024-03121-7}  \\ 

\noindent
Khan, S.A., Jagannathan, R., 1995.     
Quantum mechanics of charged particle beam transport through magnetic lenses. 
Phys. Rev. E 51, 2510-2515.  
\url{https://doi.org/10.1103/PhysRevE.51.2510}  \\    

\noindent 
Khan, S.A., Jagannathan, R., 2021.   
Quantum mechanics of round magnetic electron lenses with Glaser and power law 
models of $B(z)$.     
Optik 229, 166303. 
\url{https://doi.org/10.1016/j.ijleo.2021.166303}  \\    

\noindent 
Khan, S.A., Jagannathan, R., 2024a.   
Quantum mechanics of bending of a charged particle beam by a dipole magnet.   
In: Advances in Imaging and Electron Physics 229, 
Eds. H\"{y}tch, M., Hawkes, P.W. pp.1-41. 
(Academic Press)    
\url{https://doi.org/10.1016/bs.aiep.2024.02.001}  \\ 

\noindent
Khan, S.A., Jagannathan, R., 2024b.   
Classical and quantum mechanics of the Wien velocity filter.   
Internat. J. Theor. Phys. 63, 16. 
\url{https://doi.org/10.1007/s10773-023-05530-6}  \\   	  

\noindent
Kiesslinga, MK-H., Tahvildar-Zadehb, A.S., 2018.        
On the quantum mechanics of a single photon.  
J. Math. Phys. 59, 112302.      
\url{https://doi.org/10.1063/1.5021066}  \\ 

\noindent
Kisel, V.V., Ovsiyuk, E.M., Red'kov, V.M., Tokarevskaya, N.G., 2011.       
Maxwell equations in complex form, squaring procedure and separating the variables. 
Ricerche Mat. 60, 1-14.    
\url{https://doi.org/10.1007/s11587-010-0092-7}  \\ 

\noindent 
Korotkova, O., Testorf, M., 2023. 
Introducing JOSA A retrospectives: editorial. 
J. Opt. Soc. Am. A 40, ED3-ED4.  
\url{https://doi.org/10.1364/JOSAA.492260}  \\ 

\noindent
Koukoutsis, E., Hizanidis, K., Ram, A.K., Vahala, G., 2023.   
Dyson maps and unitary evolution for Maxwell equations in tensor dielectric media. 
Phys. Rev. A 107, 042215.    
\url{https://doi.org/10.1103/PhysRevA.107.042215}  \\ 

\noindent 
Kulyabov, D.S., 2016.     
Spinor-like Hamiltonian for Maxwellian optics.    
EPJ Web Conf. 108, 02034.     
\url{https://doi.org/10.1051/epjconf/201610802034}  \\ 

\noindent 
Kulyabov, D.S., Korolkova, A.V., Sevastianov, L.A., 2017.       
Spinor representation of Maxwell's equations.    
J. Phys: Conf. Ser. 788, 012025.          
\url{https://doi.org/10.1088/1742-6596/788/1/012025}  \\ 
	
\noindent  
Laporte, O., Uhlenbeck, G.E., 1931.     
Applications of spinor analysis to the Maxwell and Dirac equations.      
Phys. Rev. 37, 1380-1397. 
\url{http://dx.doi.org/10.1103/PhysRev.37.1380}  \\    

\noindent 
Lax, M., Louisell, W.H., McKnight, W.B., 1975. 
From Maxwell to paraxial wave optics. 
Phys. Rev. A 11, 1365-1370. 
\url{https://doi.org/10.1103/PhysRevA.11.1365}  \\ 

\noindent{Livadiotis, 2018;}
Livadiotis, G., 2018.    
Complex symmetric formulation of Maxwell equations for fields and potentials.   
Mathematics 6, 114.   
\url{https://doi.org/10.3390/math6070114}  \\ 

\noindent 
Mazharimousavi, S.H., Roozbeh, A., Halilsoy, M., 2013.    
Electromagnetic wave propagation through inhomogeneous material layers. 
J. Electromag. Waves. Appl. 27, 2065-2074.   
\url{https://doi.org/10.1080/09205071.2013.831741}  \\ 

\noindent  
Mehrafarin, M., Balajany, H., 2010. 
Paraxial spin transport using the Dirac-like paraxial wave equation.    
Phys. Lett. A 374, 1608-1610.          
\url{http://dx.doi.org/10.1016/j.physleta.2010.01.067}  \\ 

\noindent 
Mignani, R., Recami, E., Baldo, M., 1974.     
About a Dirac-like equation for the photon according to Ettore Majorana.      
Lett. Nuovo Cim. 11, 568-572. 
\url{http://dx.doi.org/10.1007/BF02812391}  \\ 

\noindent
Mingjie, L., Peng, S., Luping, D., Xiaocong, Y., 2020.       
Electronic Maxwell's equations. 
New J. Phys. 22, 113019.    
\url{https://doi.org/10.1088/1367-2630/abc853}  \\ 

\noindent  
Moli\`{e}re, G., 1950.    
Laufende elektromagnetische multipolwellen und eine neue methode der feld-quantisierung.   
Ann. der Phys. 6, 146-162.  
\url{https://doi.org/10.1002/andp.19494410119}  \\ 

\noindent 
Moses, H.E., 1959. 
Solutions of Maxwell's equations in terms of a spinor notation: The direct and inverse problem.      
Phys. Rev. 113, 1670-1679.   
\url{http://dx.doi.org/10.1103/PhysRev.113.1670}  \\ 

\noindent 
Mukunda, N., Simon, R., Sudarshan, E.C.G., 1983.     
Paraxial wave optics and relativistic front description - II The vector theory.  
Phys. Rev. A 28, 2933-2942.     
\url{https://doi.org/10.1103/PhysRevA.28.2933}  \\ 

\noindent 
Mukunda, N., Simon, R., Sudarshan, E.C.G., 1985a.     
Fourier optics for the Maxwell field: Formalism and applications.    
J. Opt. Soc. Am. A 2, 416-426.  
\url{https://doi.org/10.1364/JOSAA.2.000416}  \\ 

\noindent 
Mukunda, N., Simon, R., Sudarshan, E.C.G., 1985b.     
Paraxial Maxwell beams: Transformations by general linear optical systems.   
J. Opt. Soc. Am. A 2, 1291-1296.       
\url{https://doi.org/10.1364/JOSAA.2.001291}  \\ 
	
\noindent  
Oppenheimer, J.R., 1931. 
Note on light quanta and the electromagnetic field.   
Phys. Rev. 38, 725-747.  
\url{http://dx.doi.org/10.1103/PhysRev.38.725}  \\ 

\noindent 
Przeszowski, J.A., 2023.   
Smeared field description of free electromagnetic field. 
Acta Phys. Polon. A 143, S107-S111.   
\url{https://doi.org/10.12693/APhysPolA.143.S107}  \\ 

\noindent 
Ram, A.K., Vahala, G., Vahala, L., Soe, M., 2021.      
Reflection and transmission of electromagnetic pulses at a planar dielectric interface: Theory and quantum lattice simulations.   
AIP Advances 11, 105116.         
\url{https://doi.org/10.1063/5.0067204}  \\ 

\noindent 
Sebens, C.T., 2019.    
Electromagnetism as quantum physics.    
Found. Phys. 49, 365–389.   
\url{https://doi.org/10.1007/s10701-019-00253-3}  \\ 

\noindent 
Simon, R., Sudarshan, E.C.G, Mukunda, N., 1986.     
Gaussian Maxwell beams. 
J. Opt. Soc. Am. A 3, 536-540.   
\url{https://doi.org/10.1364/JOSAA.3.000536}  \\ 

\noindent 
Simon, R., Sudarshan, E.C.G., Mukunda, N., 1987.      
Cross polarization in laser beams.    
Appl. Opt. 26, 1589-1593.  
\url{https://doi.org/10.1364/AO.26.001589}  \\ 

\noindent 
Sudarshan, E.C.G., Simon, R., Mukunda, N., 1983.     
Paraxial wave optics and relativistic fron description - I The scalar theory. 
Phys. Rev. A 28, 2921-2932.      
\url{https://doi.org/10.1103/PhysRevA.28.2921}  \\ 
	
\noindent 
Vahala, G., Vahala, L., Soe, M., Ram, A.K., 2020a.     
Unitary quantum lattice simulations for Maxwell equations in vacuum and in dielectric media.   
J. Plasma Phys. 86, 905860518.     
\url{https://doi.org/10.1017/S0022377820001166}  \\ 
	
\noindent 
Vahala, G., Vahala, L., Soe, M., Ram, A.K., 2020b.    
Building a three-dimensional quantum lattice algorithm for Maxwell equations. 
Radiation Effects and Defects in Solids 175, 986-990.      
\url{https://doi.org/10.1080/10420150.2020.1845685}  \\ 
	
\noindent 
Vahala, G., Vahala, L., Soe, M., Ram, A.K., 2020c. 
The effect of the Pauli spin matrices on the quantum lattice algorithm for Maxwell equations in inhomogeneous media.   
arXiv:2010.12264[physics.plasm-ph].    
\url{https://arxiv.org/abs/2010.12264}  \\ 
	
\noindent 
Vahala, G., Vahala, L., Soe, M., Ram, A.K., 2021a.       
One and two-dimensional quantum lattice algorithms for Maxwell equations in inhomogeneous scalar dielectric media I: Theory.    
Radiation Effects and Defects in Solids 176, 49-63.      
\url{https://doi.org/10.1080/10420150.2021.1891058}  \\ 
	
\noindent 
Vahala, G., Vahala, L., Soe, M., Ram, A.K., 2021b.     
One and two-dimensional quantum lattice algorithms for Maxwell equations in inhomogeneous scalar dielectric media II: Simulations.    
Radiation Effects and Defects in Solids 176, 64-72.       
\url{https://doi.org/10.1080/10420150.2021.1891059}  \\ 
	
\noindent 
Vahala, G., Vahala, L., Ram, A.K., Soe, M., 2022a. 
The effect of the width of the incident pulse to the dielectric transition layer in the scattering of an electromagnetic pulse.   
arXiv:2201.09259[physics.plasm-ph]. 
\url{https://arxiv.org/abs/2201.09259}  \\ 
	
\noindent 
Vahala, G., Hawthorne, J., Vahala, L., Ram, A.K., Soe, M., 2022b.     
Quantum lattice representation for the curl equations of Maxwell equations. 
Radiation Effects and Defects in Solids 177, 85-94.     
\url{https://doi.org/10.1080/10420150.2022.2049784}  \\  

\noindent
Vahala, G., Soe, M., Koukoutsis, E., Hizanidis, K., Vahala, L., Ram, A.K., 2023. 
Qubit lattice algorithms based on the Schr\"{o}dinger-Dirac representation of Maxwell equations and their extensions. 
arXiv:2307.13182[quant-ph]. 
\url{https://arxiv.org/abs/2307.13182}  \\ 

\noindent  
Weinberger, P., 2008.     
All you need to know about the Dirac equation.  
Phil. Mag. 88, 2585-601.    
\url{https://doi.org/10.1080/14786430802247171}     

\end{document}